\documentclass[a4paper,twocolumn,11pt,accepted=2024-10-02]{quantumarticle}
\pdfoutput=1
\usepackage[utf8]{inputenc}
\usepackage{braket}
\usepackage{amsmath,amssymb,graphicx}
\usepackage{caption}
\usepackage{subcaption}
\usepackage{qcircuit}
\usepackage[dvipsnames]{xcolor}
\usepackage{verbatim}
\usepackage[normalem]{ulem}
\usepackage[stable]{footmisc}
\usepackage{graphicx}
\usepackage{todonotes}
\usepackage[rightcaption]{sidecap}
\usepackage{float}
\usepackage{multirow}

\newcommand{\norm}[1]{\left\lVert#1\right\rVert}
\newcommand{\R}{\mathbb{R}}

\def\ket#1{\mathinner{|{#1}\rangle}}

\newcommand{\revision}[1]{\textcolor{black}{#1}}
\newcommand{\revisionnew}[1]{\textcolor{black}{#1}}

\begin{document}

\title{Quantum Vision Transformers}

\author{El Amine Cherrat}
\affiliation{IRIF, CNRS - Université Paris Cité, France}

\author{Iordanis Kerenidis}
\affiliation{IRIF, CNRS - Université Paris Cité, France}
\affiliation{QC Ware, Palo Alto, USA and Paris, France}

\author{Natansh Mathur}
\affiliation{IRIF, CNRS - Université Paris Cité, France}
\affiliation{QC Ware, Palo Alto, USA and Paris, France}

\author{Jonas Landman}
\affiliation{School of Informatics, University of Edinburgh, Scotland, UK}
\affiliation{QC Ware, Palo Alto, USA and Paris, France}
\email{jonas.landman@qcware.com}
\orcid{0000-0002-2039-5308}

\author{Martin Strahm}
\affiliation{F. Hoffmann La Roche AG}

\author{Yun Yvonna Li}
\affiliation{F. Hoffmann La Roche AG}

\maketitle

\begin{abstract}
    In this work, quantum transformers are designed and analysed in detail by extending the state-of-the-art classical transformer neural network architectures known to be very performant in natural language processing and image analysis. Building upon the previous work, which uses parametrised quantum circuits for data loading and orthogonal neural layers, we introduce three types of quantum transformers for training and inference, including a quantum transformer based on compound matrices, which guarantees a theoretical advantage of the quantum attention mechanism compared to their classical counterpart both in terms of asymptotic run time and the number of model parameters. These quantum architectures can be built using shallow quantum circuits and produce qualitatively different classification models. The three proposed quantum attention layers vary on the spectrum between closely following the classical transformers and exhibiting more quantum characteristics. As building blocks of the quantum transformer, we propose a novel method for loading a matrix as quantum states as well as two new trainable quantum orthogonal layers adaptable to different levels of connectivity and quality of quantum computers. We performed extensive simulations of the quantum transformers on standard medical image datasets that showed competitively, and at times better performance compared to the classical benchmarks, including the best-in-class classical vision transformers. \revisionnew{The quantum transformers we trained on these small-scale datasets require fewer parameters compared to standard classical benchmarks. While this observation aligns with the anticipated computational benefit of our quantum attention layers, particularly regarding the size of the input images,  further validation is necessary to confirm these initial findings as quantum computers scale up.}
    Finally, we implemented our quantum transformers on superconducting quantum computers and obtained encouraging results for up to six qubit experiments.
\end{abstract}

\section{Introduction}

Quantum machine learning \cite{biamonte2017quantum} uses quantum computation in order to provide novel and powerful tools to enhance the performance of classical machine learning algorithms. Some use parametrised quantum circuits to compute quantum neural networks and explore a higher-dimensional optimisation space \cite{cong2019quantum,bharti2021noisy,cerezo2020variational}, while others exploit interesting properties native to quantum circuits, such as orthogonality or unitarity \cite{Landman2022QuantumMF, projUNNLecun}.

\begin{figure*}[]
 \centering
  \begin{minipage}{0.49\textwidth}
    \centering
    \includegraphics[height=180px]{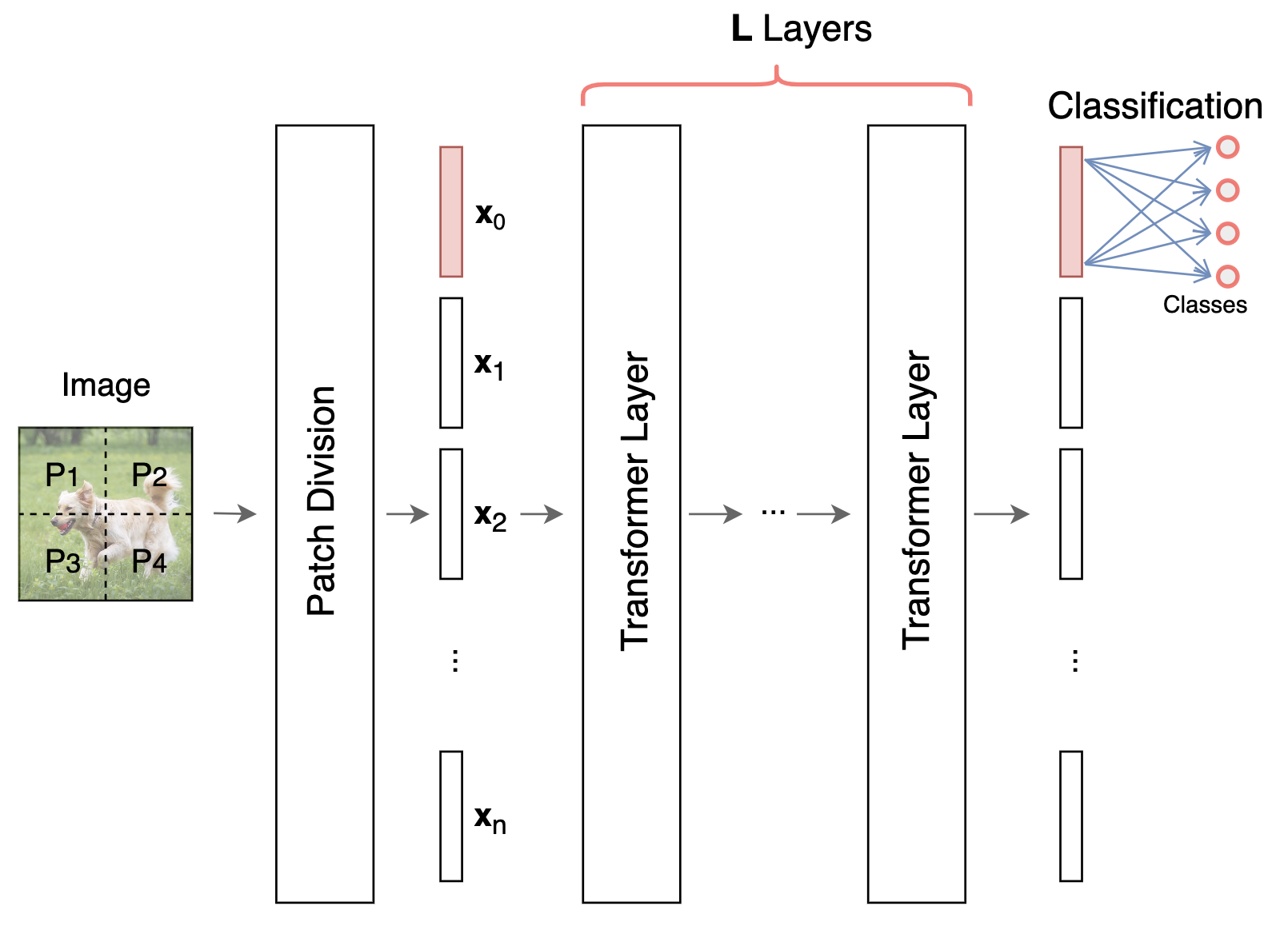}
    \caption{Vision Transformer Overview} 
    \label{fig:TransformerNetwork}
  \end{minipage}
  \begin{minipage}{0.49\textwidth}
    \centering
    \includegraphics[height=180px]{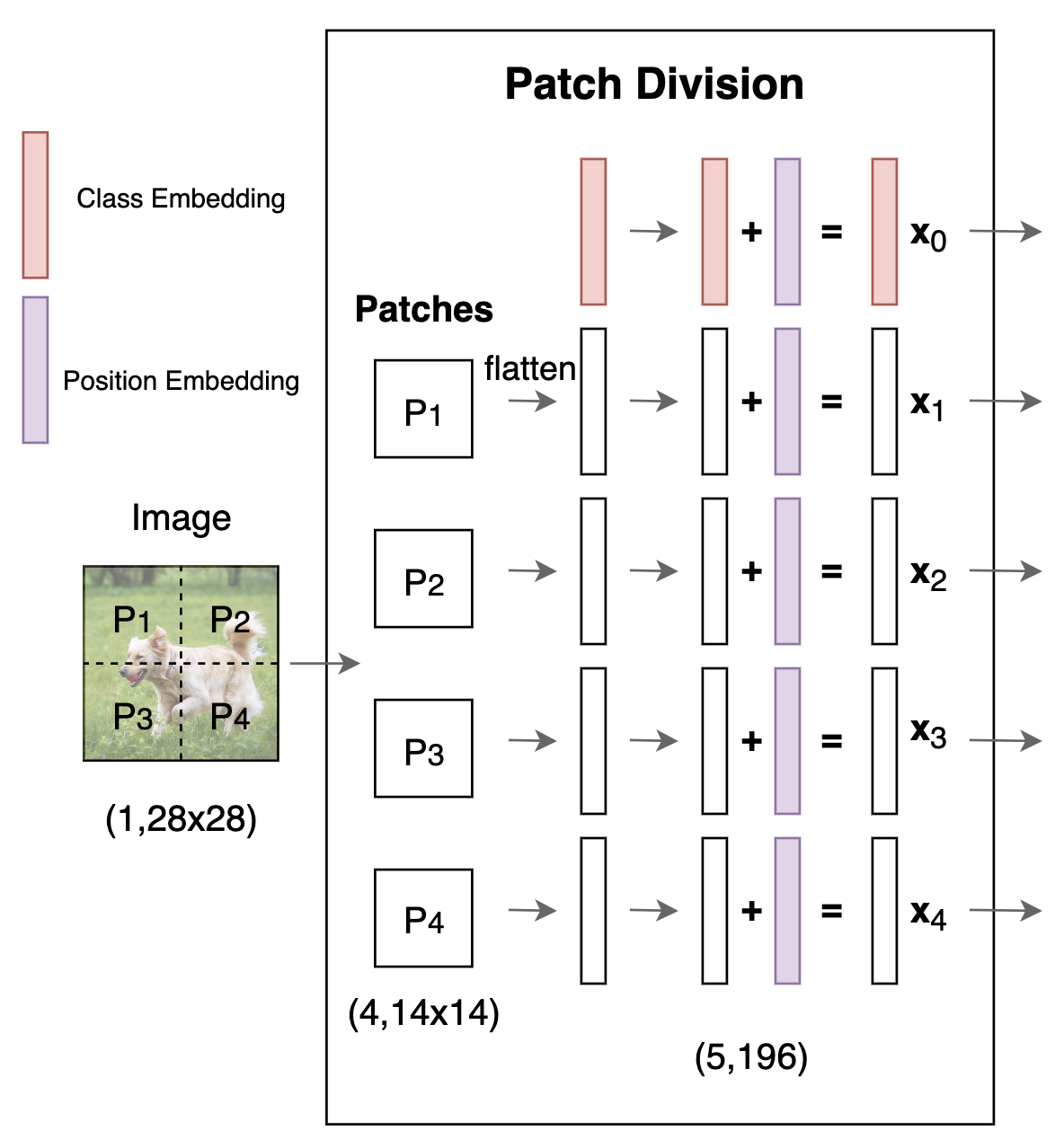}
    \caption{Patch Division Preprocessing}
    \label{fig:tranformerpatchdivision}
  \end{minipage}

    \caption*{\revision{Components of the Vision Transformer (1/2): Fig.\ref{fig:TransformerNetwork} shows the global architecture of a vision transformers. First, the image is preprocessed using patch division (Fig.\ref{fig:tranformerpatchdivision}), and then several transformer layers are applied (see details in Fig.\ref{fig:tranformerlayer} and Fig.\ref{fig:tranformerAttention}). The final step consists in a simple fully connected neural network for classification.}}
\end{figure*}

In this work, we focus on transformers, a neural network architecture proposed by \cite{attentionisallyouneed} which has been applied successfully to both natural language processing \cite{BERT} and visual tasks \cite{dosovitskiy2020image}, providing state-of-the-art performance across different tasks and datasets \cite{tay2020efficient}. \revision{While the transformer architecture and attention mechanism were notably popularized by \cite{attentionisallyouneed}, antecedents of these mechanisms can be found in earlier works. Specifically, \cite{bahdanau_neural_2016} explored such concepts in the realm of neural machine translation. In earlier works, recurrent neural network approaches hinted at the underpinnings of attention-like mechanisms, which can be found in \cite{schmidhuber_reducing_1993,schmidhuber_learning_1992}.}  At a high level, transformers are neural networks that use an \emph{attention} mechanism that takes into account the global context while processing the entire input data element-wise. For visual recognition or text understanding, the context of each element is vital, and the transformer can capture more global correlations between parts of the sentence or the image compared to convolutional neural networks without an attention mechanism \cite{dosovitskiy2020image}.
In the case of visual analysis for example, images are divided into smaller patches, and instead of simply performing patch-wise operations with fixed size kernels, a transformer learns attention coefficients per patch that weigh the attention paid to the rest of the image by each patch.

\begin{figure*}[]
  \centering

  \begin{minipage}{0.49\textwidth}
    \centering
    \includegraphics[height=170px]{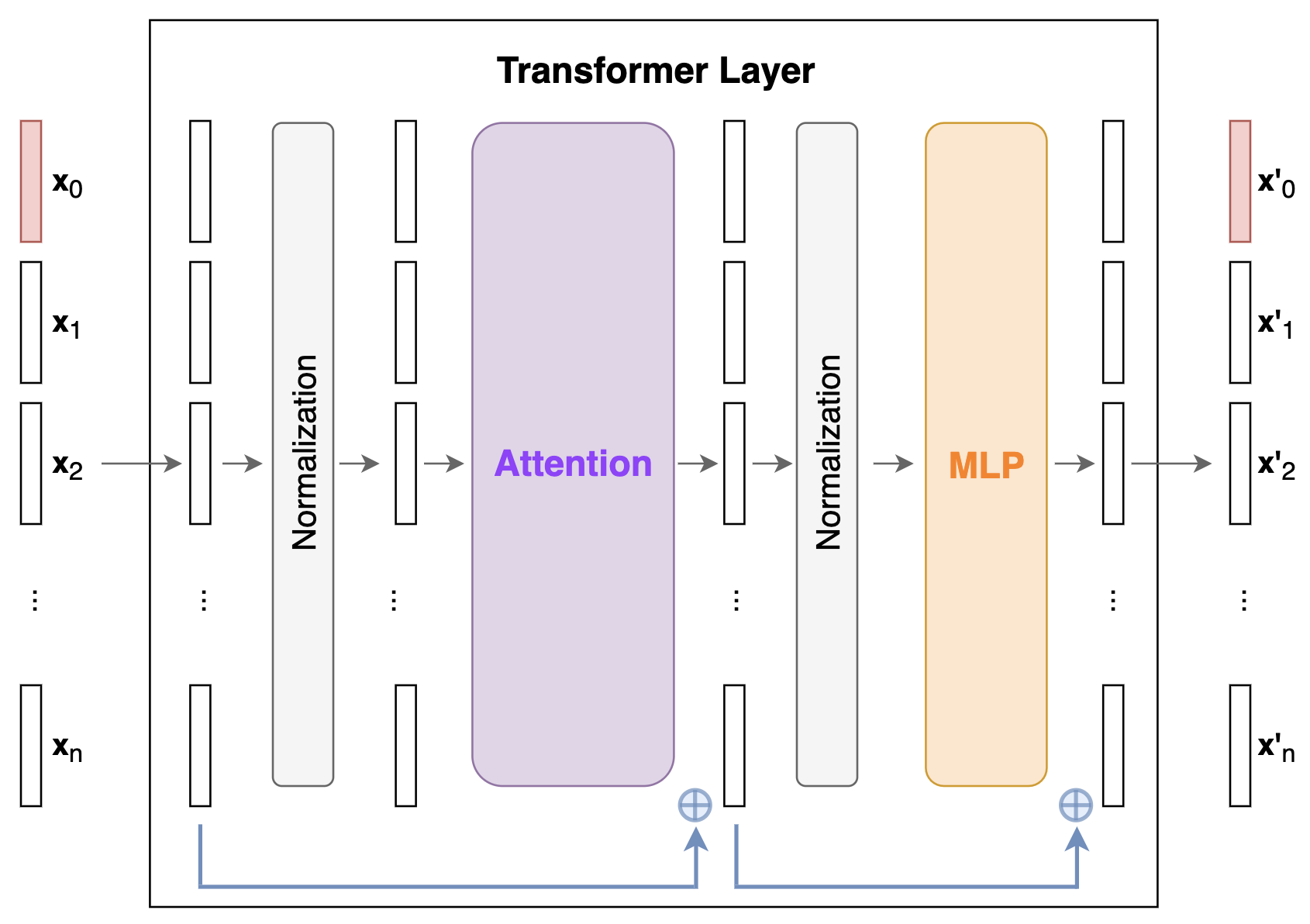}
    \caption{Transformer layer}
    \label{fig:tranformerlayer}
  \end{minipage}
  \begin{minipage}{0.49\textwidth}
    \centering
    \includegraphics[height=200px]{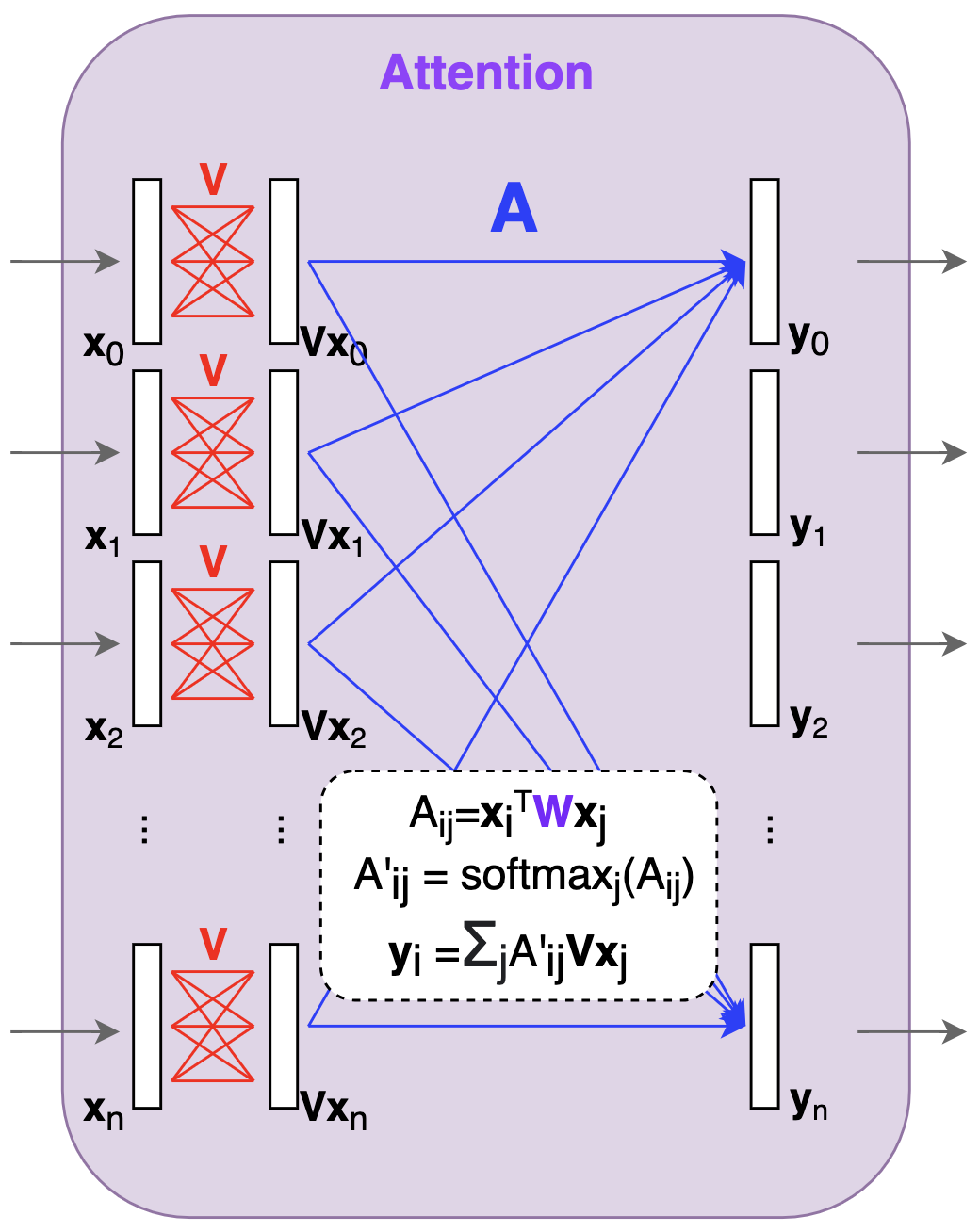}
    \caption{Attention Mechanism} 
    \label{fig:tranformerAttention}
  \end{minipage}

    \caption*{\revision{Components of the Vision Transformer (2/2): Components in a single transformer layer is outlined in (Fig.\ref{fig:tranformerlayer}). At its core, the attention mechanism learns how to weigh different parts of the input (Fig.\ref{fig:tranformerAttention}),  where the trainable matrices are denoted by $\mathbf{V}$ and $\mathbf{W}$. This attention mechanism is the focus of our quantum circuits.}}
\end{figure*}

In one related work, classical transformer architectures and attention mechanisms have been used to perform quantum tomography \cite{cha2021attention}. Moreover, a quantum-enhanced transformer for sentiment analysis has been proposed in \cite{di2022dawn}, and a \emph{self-attention} mechanism for text classification has been used in \cite{li2022quantum}. These works use standard variational quantum circuits to compute the neural networks, and the attention coefficients are calculated classically. A method for using a natively quantum attention mechanism for reinforcement learning has also been proposed in \cite{sanches2022short}. \cite{yang2022semiconductor} performed semiconductor defect detection using quantum self-attention, also using standard variational quantum circuits. We also note the proposals of \cite{cong2019quantum, henderson2020quanvolutional} for variational circuits with similarities to convolutional neural networks for general purpose image classification.

The difference between the above-mentioned approaches and the proposed approached of this work mainly stems from the linear algebraic tools we developed which make our quantum circuits much more Noisy Intermediate-Scale Quantum (NISQ)-friendly with proven scalability in terms of run time and model parameters, in contrast to variational quantum circuit approaches taken in \cite{QNN2018, cerezo2020variational}which lack proof of scalability \cite{mitarai2018quantum}. This advantage in scalability of our proposed parametrised quantum circuits is made possible by the use of a specific \emph{amplitude encoding} for translating vectors as quantum states, and consistent use of hamming-weight preserving quantum gates instead of general quantum ansatz. In addition to a quantum translation of the classical vision transformer, a novel and natively quantum method is proposed in this work, namely the {\em compound transformer}, which invokes Clifford Algebra operations that is hard to compute classically.

While we adapted the vision transformer architecture to ease the translation of the attention layer into quantum circuits and benchmarked our methods on vision tasks, the proposed approaches for quantum attention mechanism can be easily adapted to apply to other fields of applications, for example in natural language processing where transformers have been proven to be particularly efficient \cite{BERT}.

\begin{figure*}[]
  \centering
  \includegraphics[width=0.9\textwidth]{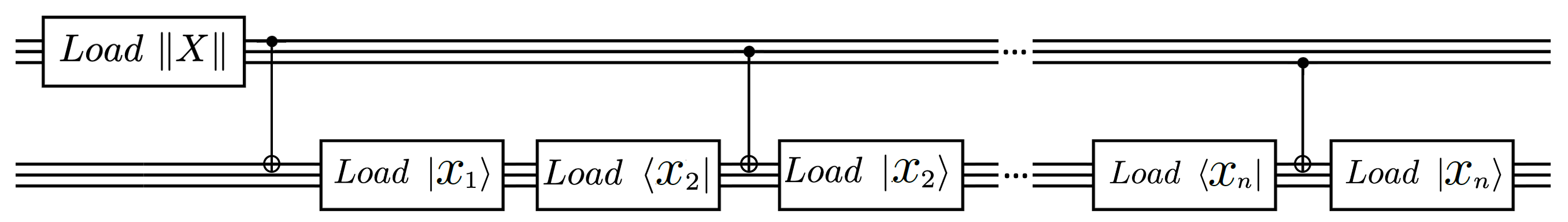}
  \caption{Data loader circuit for a matrix $X\in\R^{n\times d}$. The top register uses $N$ qubits and the vector data loader to load the norms of each row, $(\norm{\mathbf{x}_1},\cdots,\norm{\mathbf{x}_n})$, to obtain the state $\frac{1}{\norm{\mathbf{X}}}\sum_{i=1}^n \norm{\mathbf{x}_i}\ket{\mathbf{e}_i}$. The lower register uses $d$ qubits to load each row $\mathbf{x}_i \in \R^d$ sequentially, by applying the vector loader and their adjoint for each row $\mathbf{x}_i$, with CNOTs controlled by the corresponding qubit $i$ of the top register. Each loader on the lower register has depth $\mathcal{O}(\log d)$.
s  }
  \label{fig:dataloader_matrix}
\end{figure*}

The main ingredient in a transformer as introduced by \cite{dosovitskiy2020image} is the \emph{attention layer}, shown in Fig.\ref{fig:tranformerAttention}. This attention layer is also the focus of this work which seeks to leverage quantum circuit for computational advantages. Given an input image $\mathbf{X}\in\mathbb{R}^{n\times d}$, we transform the input data into $n$ patches each with dimension of $d$, and denote each patch $i$ with $\mathbf{x}_i \in \R^d$. The trainable weight matrix from the linear fully connected layer at the beginning of each attention layer is denoted by $\mathbf{V}$. The heart of the attention mechanism, i.e. the attention coefficients which weighs each patch $\mathbf{x}_i$ to every other patch is denoted by: $$A_{ij} = \mathbf{x}_i^T \mathbf{W} \mathbf{x}_j,$$ where $\mathbf{W}$ represents the second trainable weight matrix.

Based on the architecture shown in Fig.\ref{fig:tranformerAttention} we propose three types of quantum transformers (Sections \ref{sec:ortho_patch_wise}, \ref{sec:ortho_attention} and \ref{sec:compound_transformers}) and apply these novel architectures to visual tasks for benchmarking. Section \ref{sec:quantum_direct_attention} outlines the approach of combining \ref{sec:ortho_patch_wise} and \ref{sec:ortho_attention} into one circuit to perform inference on the quantum circuit once the attention coefficients have been trained, while sections  \ref{sec:ortho_patch_wise}, \ref{sec:ortho_attention} and \ref{sec:compound_transformers} propose 3 distinct quantum architecture for training and inference.

The first quantum transformer introduced in Section \ref{sec:ortho_patch_wise} implements a trivial attention mechanism which where each patch pays attention only to itself while retaining the beneficial property of guaranteed orthogonality of trained weight matrices \cite{jia2019orthogonal}. In the second quantum transformer introduced in Section \ref{sec:ortho_attention}, coined the Orthogonal Transformer, we design a quantum analogue for each of the two main components of a classical attention layer: a linear fully connected layer and the attention matrix to capture the interaction between patches. This approach follows the classical approach quite closely.
In Section \ref{sec:compound_transformers}, the Compound Transformer, which takes advantage of the quantum computer to load input states in superposition, is defined. For each of our quantum methods, we provide theoretical analysis of the computational complexity of the quantum attention mechanisms which is lower compared to their classical counterparts.

The mathematical formalism behind the Compound Transformer is the second-order compound matrix \cite{horn2012matrix}. Compound Transformer uses quantum layers to first load all patches into the quantum circuit in uniform superposition and then apply a single unitary to multiply the input vector in superposition with a trainable second-order compound matrix \cite{subspace_states_qcware}. Here both the input vector and the trainable weight matrix are no longer a simple vector or a simple matrix. Details are given in Sections \ref{sec:quantum_transformers} and  \ref{sec:compound_transformers}.

The fundamental building blocks for the implementation of a transformer architecture including the matrix data loader and quantum orthogonal layers are introduced in Sections \ref{sec:quantum_data_loaders}, \ref{sec:quantum_ortho_layers}.

\section{Quantum Tools}

In this work, we will use the RBS gate given in Eq.(\ref{RBS}). RBS gates implement the following unitary:
\begin{equation} \label{RBS}
  \text{RBS}(\theta) = \left( \begin{array}{cccc}
      1 & 0             & 0            & 0 \\
      0 & \cos(\theta)  & \sin(\theta) & 0 \\
      0 & -\sin(\theta) & \cos(\theta) & 0 \\
      0 & 0             & 0            & 1\end{array} \right)
\end{equation}

This gate can be implemented rather easily, either as a native gate, known as FSIM \cite{foxen2020demonstrating}, or using four Hadamard gates, two $R_y$ rotation gates, and two two-qubits CZ gates:

\begin{figure}[h]
  \centering
  \includegraphics[width=120px]{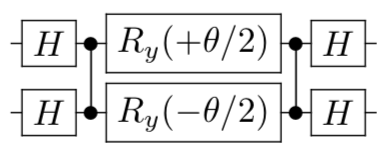}
  \caption{A possible decomposition of the $RBS(\theta)$ gate.}
  \label{fig:RBS_implementation}
\end{figure}

\subsection{Quantum Data Loaders for Matrices\footnote{For this section, details are provided in \ref{sec:quantum_data_loaders_long}.}}
\label{sec:quantum_data_loaders}

Loading a whole matrix $\mathbf{X} \in \R^{n\times d}$ in a quantum state is a powerful technique for machine learning. \cite{NearestCentroid2021} designed quantum circuits to load input vectors using \revision{using $N=n+d$ qubits with a} \emph{unary} amplitude encoding, more specifically a basis of states of hamming weight 1 where all qubits are in state 0 except one in state 1 is used. The number of required gates to load a vector is $d-1$.
In this work, we extend their approach to build a data loader for matrices (Fig.\ref{fig:dataloader_matrix}) where every row of $\mathbf{X}$ is loaded in superposition. The required number of gates to load a matrix is $(n-1) + (2n-1)(d-1)$.
The resulting state of the matrix loader shown in Fig.\ref{fig:dataloader_matrix} is a superposition of the form:

\begin{equation}
  \ket{\mathbf{X}} = \frac{1}{\norm{\mathbf{X}}}\sum_{i=1}^n\sum_{j=1}^d X_{ij}\ket{\mathbf{e}_j}\ket{\mathbf{e}_i}
\end{equation}

\subsection{Quantum Orthogonal Layers\footnote{Refer to \ref{sec:quantum_ortho_layers_long} for additional details about this section.}}
\label{sec:quantum_ortho_layers}

The classical attention layer (Fig.\ref{fig:tranformerAttention}) starts with a linear fully connected layer, where each input, i.e. patch $\mathbf{x}_i$, is a vector and is multiplied by a weight matrix $\mathbf{V}$. To perform this operation quantumly we generalise the work of \cite{Landman2022QuantumMF}, where a quantum orthogonal layer is defined as a quantum circuit applied on a state $\ket{\mathbf{x}}$ (encoded in the \emph{unary} basis) to produce the output state $\ket{\mathbf{V}\mathbf{x}}$. More precisely, $\mathbf{V}$ is the matrix corresponding to the unitary of the quantum layer, restricted to the \emph{unary} basis. \revision{This matrix is orthogonal due to the unitary nature of quantum operations.}

\begin{figure*}[]
    \centering
    \begin{minipage}{0.39\textwidth}
        \centering
        \includegraphics[height=8em]{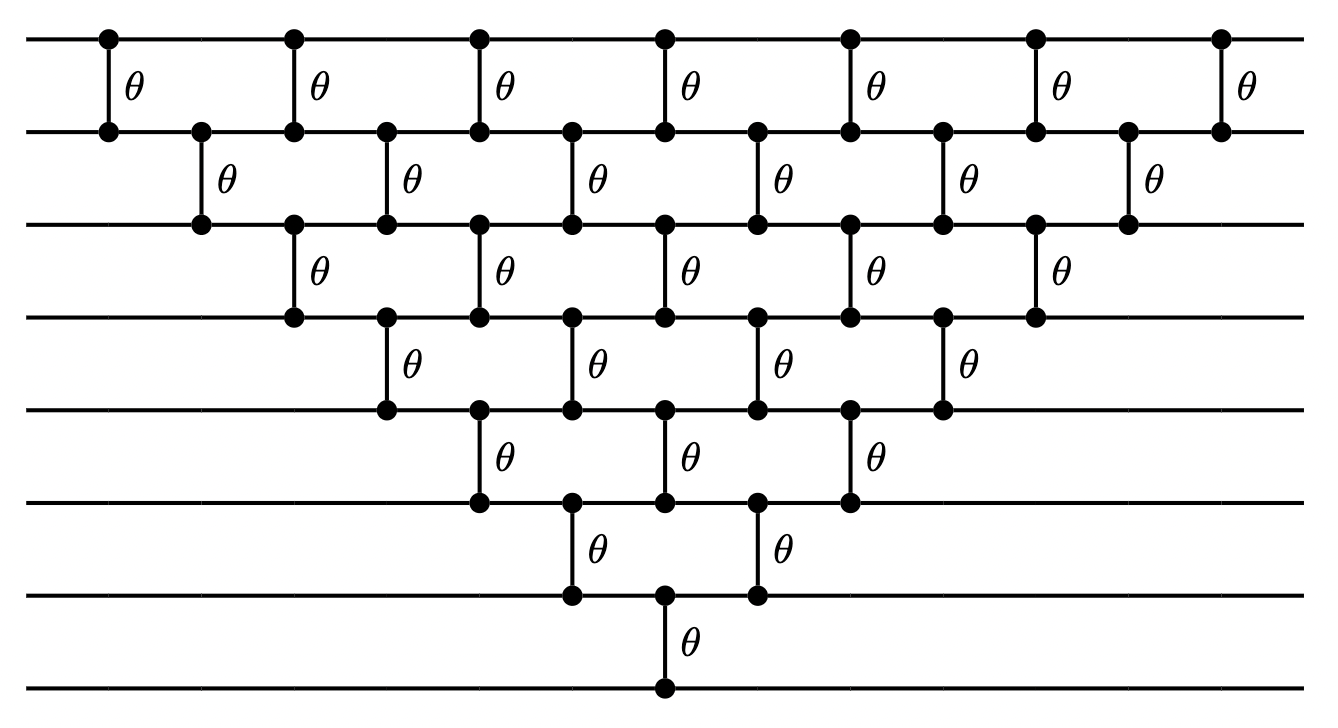}
        \caption{Pyramid Circuit}
        \label{fig:pyramid}
    \end{minipage}
    \begin{minipage}{0.29\textwidth}
        \centering
        \includegraphics[height=8em]{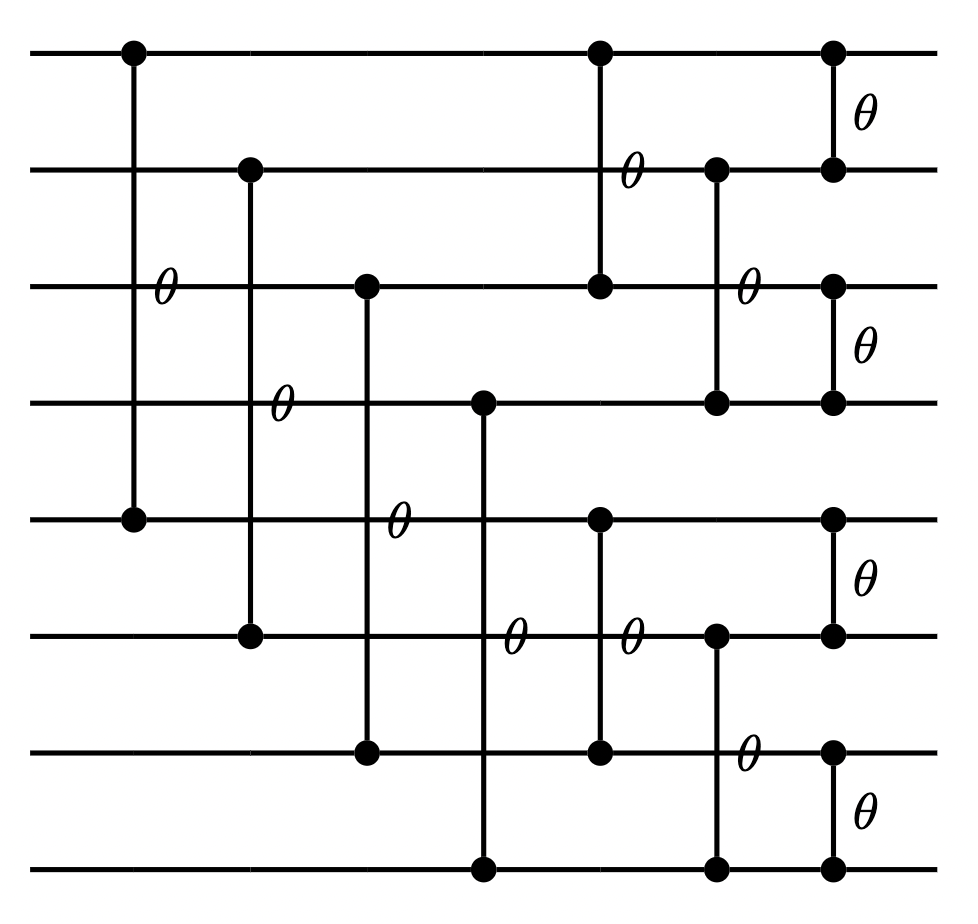} 
        \caption{Butterfly Circuit}
        \label{fig:butterfly}
    \end{minipage}
    \begin{minipage}{0.29\textwidth}
        \centering
        \includegraphics[height=7.5em]{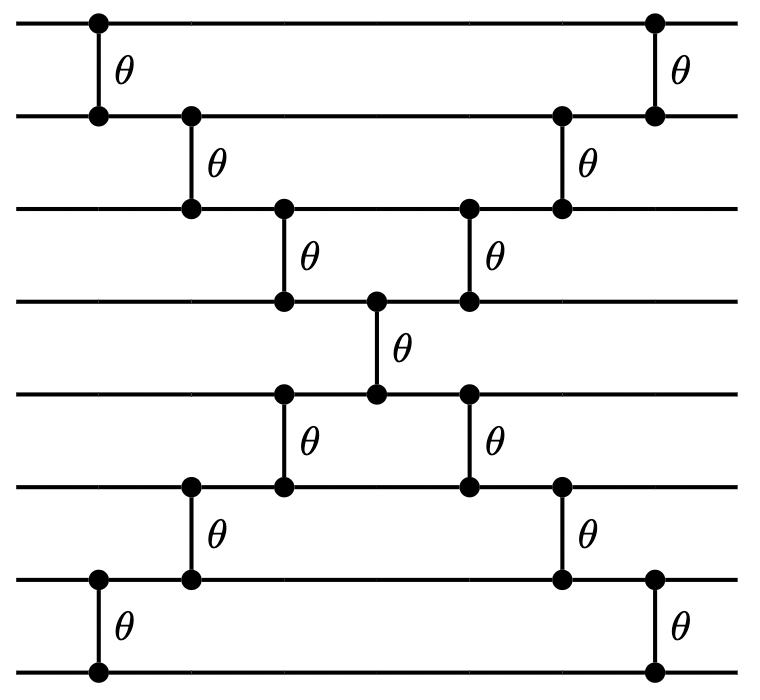} 
        \caption{X Circuit}
        \label{fig:Xcircuit}
    \end{minipage}  
\caption*{Quantum Orthogonal Layers. Vertical lines represent two-qubit RBS gates, parametrized with independent angles $\theta$.}
\end{figure*}

In addition to the already existing Pyramid circuit (Fig.\ref{fig:pyramid}) from \cite{Landman2022QuantumMF}, we define two new types of quantum orthogonal layers with different levels of expressivity and resource requirements: the butterfly circuit (Fig.\ref{fig:butterfly}), 
and the $X$ circuit (Fig.\ref{fig:Xcircuit}). 

\begin{table}[]
  \centering
  \resizebox{0.5\textwidth}{!}{
    \begin{tabular}{|c|c|c|c|}
      \hline
      Circuit   & Hardware Connectivity        & Depth     & \# Gates              \\
      \hline
      Pyramid   & \revision{Nearest Neighbour} & $2N-3$    & $\frac{N(N-1)}{2}$    \\
      X         & \revision{Nearest Neighbour} & $N-1$     & $2N-3$                \\
      Butterfly & All-to-all                   & $\log(N)$ & $\frac{N}{2} \log(N)$ \\
      \hline
    \end{tabular}
  }
  \caption{Comparison of different quantum orthogonal layer circuits with $N$ qubits.}
  \label{table:circuit_comparison}
\end{table}

Looking at Table \ref{table:circuit_comparison}, the $X$ circuit is the most suited for noisy hardware. It requires smaller number of gates while maintaining a path from every input qubit to every output qubit. It is also less expressive with a restrained set of possible orthogonal matrices and fewer trainable parameters. The butterfly circuit requires logarithmic circuit depth, a linear number of gates, and exhibits a higher level of expressivity. It originates from the classical Cooley–Tukey algorithm \cite{cooley1965algorithm} used for Fast Fourier Transform \revision{and, it performs an operation analogous to the method presented in \cite{Jing2016TunableEU} for classical recurrent neural networks when it is implemented with RBS gates}. Note that the butterfly circuit requires the ability to apply gates on all possible qubit pairs.\\

As shown in \cite{subspace_states_qcware}, quantum orthogonal layers can be generalised to work with inputs which encode a vector on a larger basis. 
Namely, instead of the \emph{unary} basis, where all qubits except one are in state $0$, basis of hamming weight $k$ can be used as well. A basis of hamming weight $k$ comprises of $N \choose k$ possible states over $N$ qubits. A vector $\mathbf{x}\in\R^{N \choose k}$ can be loaded as a quantum state $\ket{\mathbf{x}}$ using only $N$ qubits.
Since the quantum orthogonal layers are hamming weight preserving circuits, the output state from such circuits will also be a vector encoded in the same basis. 
Let $\mathbf{V}$ be the matrix corresponding to the quantum orthogonal layer in the \emph{unary} basis, and $\mathbf{x}$ of hamming weight $k$, the output state will no longer be $\ket{\mathbf{Vx}}$, but instead $\ket{\boldsymbol{\mathcal{V}}^{(k)}\mathbf{x}}$, where $\boldsymbol{\mathcal{V}}^{(k)}$ is the \emph{$k$-th order compound matrix} of $\mathbf{V}$ \cite{horn2012matrix}. We can see $\boldsymbol{\mathcal{V}}^{(k)}$ as the expansion of $\mathbf{V}$ in the hamming weight k basis.
More precisely, given a matrix $\mathbf{V} \in \R^{N\times N}$, the $k^{th}$-order compound matrix $\boldsymbol{\mathcal{V}}^{(k)}$ for $k \in [N]$ is the ${N}\choose{k}$ dimensional matrix with entries: $$\mathcal{V}^{(k)}_{IJ} = \text{det}(\mathbf{V}_{IJ}),$$ where $I$ and $J$ are subsets of rows and columns of $\mathbf{V}$ with size k. 

\revision{Recent research supports the trainability of the quantum layers presented in this paper. \cite{monbroussou2023trainability} provide evidence for the trainability and expressivity of hamming weight preserving circuits, indicating that our layers are not prone to the vanishing gradients problem, commonly referred to as barren plateaus. This assertion is further reinforced by studies in \cite{fontana2023adjoint, ragone2023unified}. Nonetheless, the existence and implications of exponential local minima \cite{You2021ExponentiallyML, Anschuetz2022QuantumVA} within our framework remain an open question.}

\section{Quantum Transformers}
\label{sec:quantum_transformers}

The second component of the classical attention layer is the interaction between patches (Fig.\ref{fig:tranformerAttention}) where the attention coefficients $A_{ij} = \mathbf{x}_i^T \mathbf{W} \mathbf{x}_j $ is trained by performing $\mathbf{x}_i^T\mathbf{W} \mathbf{x}_j$ for a trainable orthogonal matrix $\mathbf{W}$ and all pairs of patches $\mathbf{x}_i$ and $\mathbf{x}_j$. 
After that, a non-linearity, for example softmax, is applied to obtain each output $\mathbf{y}_i$. 
Three different approaches for implementing the quantum attention layer are introduced in the next sections, listed in the order of increasing complexity in terms of quantum resource requirement, which reflect the degree to which quantum circuits are leveraged to replace the attention layer.  
A comparison between these different quantum methods is provided in Table \ref{table:quantum_transformer_comparison}, which is applicable to both training and inference.

Table \ref{table:quantum_transformer_comparison} lists 5 key parameters of the proposed quantum architecture which reflect their theoretical scalability. The number of trainable parameters for a classical vision transformer is $2d^2$ (see  Section \ref{sec:classical_transformer}), which can be directly compared with the number of trainable parameters of the proposed quantum approaches. The number of fixed parameters per quantum architecture is required for data loading. \revision{In this table, the circuit depth represents the combined depth of both the data loader and the quantum layer. Furthermore, the butterfly layer detailed in Fig.\ref{fig:butterfly} and the diagonal data-loader illustrated in Fig.\ref{fig:dataloaders_vector} is employed, which adds logarithmic depth for loading each vector.} The circuit depth together with the number of distinct circuits dictate the overall run time of the quantum architectures, which
can be compared to the run time of the classical transformer of $ \mathcal{O}(nd^2 + n^2d)$ (listed under the column \emph{Circuit Depth}). The number of distinct circuits per quantum architecture indicate the possibility for each architecture to be processed in parallel, akin to multi-core CPU processing. 

\begin{table*}[]
\centering
\resizebox{\linewidth}{!}{
\begin{tabular}{|l|ccccc|}
\hline
\revision{Transformer architecture}                          & \# Qubits & Circuit depth & \# Trainable parameters & \# Fixed parameters & \#  Distinct circuits \\ \hline 
A - Orthogonal Patch-wise          & $d$       & $\mathcal{O}(\log d)$                              & $\mathcal{O}(d\log d)$           & $d-1$               & $n$                    \\
B - Quantum Orthogonal Transformer & $d$       & $\mathcal{O}(\log d)$                              & $\mathcal{O}(d\log d)$           & $3(d-1)$            & $n+n^2$                  \\
C - Quantum Attention Mechanism    & $n+d$     & $\mathcal{O}(\log n + n \log d + \log d)$          & $\mathcal{O}(d\log d)$           & $n-1+(2n-1)(d-1)$   & $n$                        \\
D - Compound Transformer           & $n+d$     & $\mathcal{O}(\log n + n \log d + \log (n+d))$      & $\mathcal{O}((n+d)\log(n+d)))$   & $n-1+(2n-1)(d-1)$   & 1                          \\
Classical Transformer              & -         & $\mathcal{O}(nd^2+n^2d)$                           & $\mathcal{O}(2d^2)$              & -   & -                     \\ \hline
\end{tabular}
}
\caption{Comparison of different quantum methods to perform a single attention layer of a transformer network. $n$ and $d$ stand respectively for the number of patches and their individual dimension. All quantum orthogonal layers are implemented using the butterfly circuits. See Section \ref{sec:quantum_transformers} for details.}
\label{table:quantum_transformer_comparison}
\end{table*}

\subsection{Orthogonal Patch-wise Neural Network}
\label{sec:ortho_patch_wise}

\begin{figure}[h]
    \centering
    \includegraphics[width=0.3\textwidth]{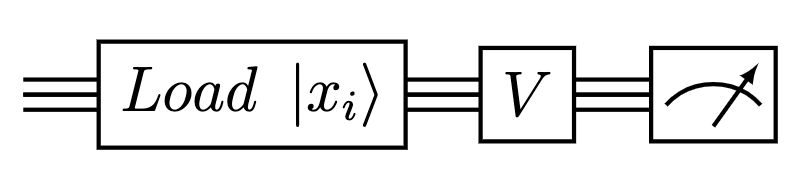}
    \caption{Quantum circuit to perform the matrix multiplication $\mathbf{V}\mathbf{x}_i$ (fully connected layer) using a data loader for $\mathbf{x}_i$ and a quantum orthogonal layer for $\mathbf{V}$.}
        \label{fig:patchwise_ortho_tranformers}
\end{figure}

The \emph{orthogonal patch-wise neural network} can be thought of as a transformer with a trivial attention mechanism, where each patch pays attention only to itself. As illustrated in Fig \ref{fig:patchwise_ortho_tranformers}, each input patch is multiplied by the same trainable matrix $\mathbf{V}$ and one circuit per patch is used. Each circuit has \revision{$N=d$ qubits} and each patch $\mathbf{x}_i$ is encoded in a quantum state with a vector data loader. A quantum orthogonal layer is used to perform multiplication of each patch with $\mathbf{V}$. The output of each circuit is a quantum state encoding $\mathbf{V} \mathbf{x}_i$, a vector which is retrieved through tomography. \revision{Importantly, this tomography procedure deals with states of linear size in relation to the number of qubits, avoiding the exponential complexity often associated with quantum tomography.}

The computational complexity of this circuit is calculated as follows: from Section \ref{sec:quantum_data_loaders}, a data loader with \revision{$N=d$ qubits} qubits has a complexity of $\log(d)$ steps. For the orthogonal quantum layer, as shown in Table \ref{table:circuit_comparison}, a butterfly circuit takes $\log(d)$ steps, with $\frac{d}{2}\log(d)$ trainable parameters. Overall, the complexity is $\mathcal{O}(\log(d))$ and the trainable parameters are $\mathcal{O}(d \log d)$. Since this circuit uses one vector data loader, the number of fixed parameters required is $d-1$.

\subsection{Quantum Orthogonal Transformer}
\label{sec:ortho_attention}

\begin{figure}[!h]
    \centering
    \includegraphics[width=0.4\textwidth]{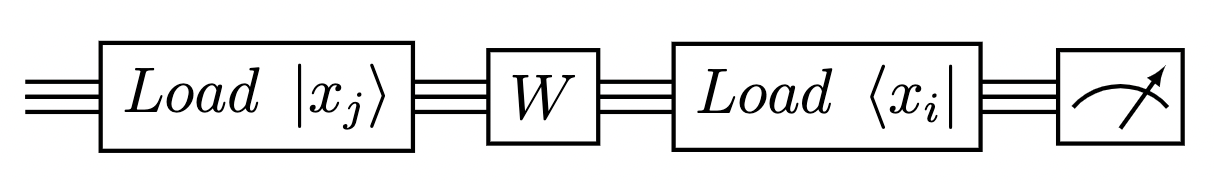}
    \caption{Quantum circuit to compute $|\mathbf{x}_i^T\mathbf{W}\mathbf{x}_j|^2$, a single attention coefficient, using data loaders for $\mathbf{x}_i$ and $\mathbf{x}_j$ and a quantum orthogonal layer for $\mathbf{W}$.}
    \label{fig:quantum_attention_circuit}
\end{figure}

Looking at Fig.\ref{fig:quantum_attention_circuit}, each attention coefficient $A_{ij} = \mathbf{x}_i ^T\mathbf{W} \mathbf{x}_j$, $\mathbf{x}_j$ is calculated first by loading $\mathbf{x}_j$ into the circuit with a vector loader followed by a trainable quantum orthogonal layer, $\mathbf{W}$, resulting in the vector $\mathbf{W} \mathbf{x}_j$.
Next, an inverse data loader of $\mathbf{x}_i$ is applied, creating a state where the probability of measuring $1$ on the first qubit is exactly $|\mathbf{x}_i ^T\mathbf{W} \mathbf{x}_j|^2 = A_{ij}^2$. 

Note the square that appears in the quantum circuit is already one type of non-linearity. Using this method, coefficients of $\mathbf{A}$ are always positive, which can still be learned during training as we show later in the Section \ref{sec:experiments}. Additional methods also exist to obtain the sign of the inner product \cite{Landman2022QuantumMF}. The estimation of $A_{ij}$ (and therefore $A'_{ij}$ if needed, by applying a column-wise \emph{softmax} classically) is repeated for each pair of patches and the same trainable quantum orthogonal layer $\mathbf{W}$. The computational complexity of this quantum circuit is similar to the previous one, with one more data loader.

Putting Figures \ref{fig:patchwise_ortho_tranformers} and \ref{fig:quantum_attention_circuit} together: the quantum circuit presented in Section \ref{sec:ortho_patch_wise} is implemented to obtain each $\mathbf{V} \mathbf{x}_j$. At the same time, each attention coefficient $|\mathbf{x}_i ^T\mathbf{W} \mathbf{x}_j|^2$ is computed on the quantum circuit, which is further post-processed column-wise with the softmax function to obtain the $A'_{ij}$. The two parts can then be classically combined to compute each $\mathbf{y}_i = \sum_j A'_{ij}\mathbf{V}\mathbf{x}_j$. In this approach, the attention mechanism is implemented by using hamming weight preserving parametrised quantum circuits to compute the weight matrices $\mathbf{V}$ and $\mathbf{W}$ separately. For computing $|\mathbf{x}_i ^T\mathbf{W} \mathbf{x}_j|^2$, we would require two data loaders ($2\times (d-1)$ gates) for $\mathbf{x}_i$ and $\mathbf{x}_j$, and one Quantum Orthogonal Layer ($d\log d$ gates in the case of Butterfly layer) for $\mathbf{W}$. To obtain $\mathbf{V}\mathbf{x}_j$, we require $d-1$ gates to load each $\mathbf{x}_j$ and a Quantum Orthogonal Layer ($d\log d$ gates in the case of Butterfly layer) for the matrix $\mathbf{V}$.

\subsection{Direct Quantum Attention}
\label{sec:quantum_direct_attention}

\begin{figure}[h]
    \centering
    \includegraphics[width=0.4\textwidth]{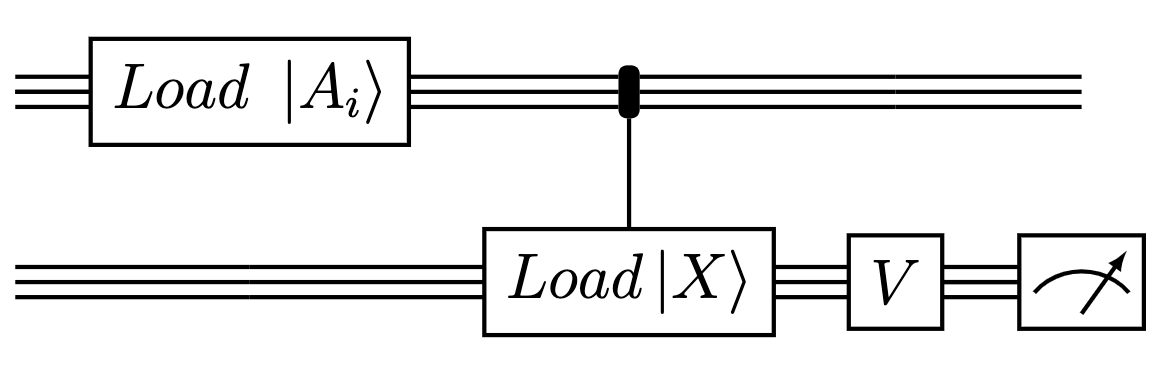}
    \caption{Quantum circuit to directly apply the attention mechanism, given each coefficient in $\mathbf{A}$. The first part of the circuit corresponds to the matrix data loader from Fig.\ref{fig:dataloader_matrix}, where $Load(\norm{\mathbf{X}})$ is replaced by $Load(\mathbf{A}_i)$. A quantum orthogonal layer from Section \ref{sec:quantum_ortho_layers} is used for $\mathbf{V}$.}
    \label{fig:direct_attention_quantum}
\end{figure}

In Section \ref{sec:ortho_attention}, the output of the attention layer $\mathbf{y}_i = \sum_j A'_{ij}\mathbf{V}\mathbf{x}_j$ is computed classically once the quantities $A'_{ij}$ and $\mathbf{V}\mathbf{x}_j$ have been computed separately with the help of quantum circuits. During inference, where the matrices $\mathbf{V}$ and $\mathbf{W}$ have been learnt, and the attention matrix $\mathbf{A}$ (or $\mathbf{A}'$) is stored classically, \emph{Direct Quantum Attention} implements the attention layer directly on the quantum computer. The matrix data loader from Fig.\ref{fig:dataloader_matrix} is used to compute each $\mathbf{y}_i = \sum_j A_{ij}\mathbf{V}\mathbf{x}_j$ using a single quantum circuit.

In Fig.\ref{fig:direct_attention_quantum}, $\mathbf{y}_i$, which corresponds to the output patch with index $i$, is computed using a quantum circuit \revision{using $N=n+d$ qubits}. These qubits are split into two main registers. 
On the top register \revision{($n$ qubits)}, the vector $\mathbf{A}_i$, $i^{th}$ row of the attention matrix $\mathbf{A}$ (or $\mathbf{A}'$), is loaded via a vector data loader, as 
$
    \sum_j A_{ij}\ket{\mathbf{e}_j}\ket{0}
$.

Next, on the lower register \revision{($d$ qubits)}, as in Fig.\ref{fig:dataloader_matrix}, the data loader for each vector $\mathbf{x}_i$, and their respective adjoint, are applied sequentially, with CNOTs controlled on each qubit $i$ of the top register. This gives the quantum state $\sum_j A_{ij} \ket{\mathbf{e}_j}\ket{\mathbf{x}_j}$, i.e. the matrix $\mathbf{X}$ is loaded with all rows re-scaled according to the attention coefficients. As for any matrix data loader, this requires $(n-1) + (2n-1)(d-1)$ gates with fixed (non trainable) parameters.

The last step consists of applying the quantum orthogonal layer $\mathbf{V}$ that has been trained before on the second register of the circuit. As previously established, this operation performs matrix multiplication between $\mathbf{V}$ and the vector encoded on the second register. Since the $k^{th}$ element of the vector $V x_j$ can be written as $\sum_q V_{kq}X_{jq}$, we get:

\begin{align}
  \label{emc}
    \sum_j A_{ij} \ket{\mathbf{e}_j}\ket{\mathbf{V}\mathbf{x}_j}\quad\quad\quad\quad\quad\quad\quad\quad \nonumber \\ 
    = \sum_j A_{ij} \ket{\mathbf{e}_j} \sum_k (\sum_q V_{kq}X_{jq})\ket{\mathbf{e}_k} \nonumber \\ 
    = \sum_k \sum_j A_{ij} (\sum_q V_{kq}X_{jq}) \ket{\mathbf{e}_j}\ket{\mathbf{e}_k} 
\end{align}

Since $\mathbf{y}_i =  \sum_j A_{ij} \mathbf{V}\mathbf{x}_j$, its $k^{th}$ element can be written $y_{ik} = \sum_j A_{ij} (\sum_q V_{kq}X_{jq})$. Therefore, the quantum state at the end of the circuit can be written as $\ket{\mathbf{y}_i} = \sum_k y_{ik}\ket{\phi_k} \ket{\mathbf{e}_k}$ for some normalised states $\ket{\phi_k}$. Performing tomography on the second register generates the output vector $\mathbf{y}_i$. 

This circuit is a more direct method to compute each $\mathbf{y}_i$. Each $\mathbf{y}_i$ uses a different $A_i$ in the first part of the circuit. 
As shown in Table \ref{table:quantum_transformer_comparison}, compared with the previous method, this method requires fewer circuits to run, but each circuit requires more qubits and a deeper circuit.
To analyse the computational complexity: the first data loader on the top register has $n$ qubit and $ \log n$ depth; the following $2n-1$ loaders on the bottom register have $d$ qubits, so $(2n-1) \log d$ depth; and the final quantum orthogonal layer $\mathbf{V}$ implemented using a butterfly circuit, has a depth of $\log d$ and $\mathcal{O}(d \log d)$ trainable parameters. 

\subsection{Quantum Compound Transformer}
\label{sec:compound_transformers}

\begin{figure}[h]
    \centering
    \includegraphics[width=0.3\textwidth]{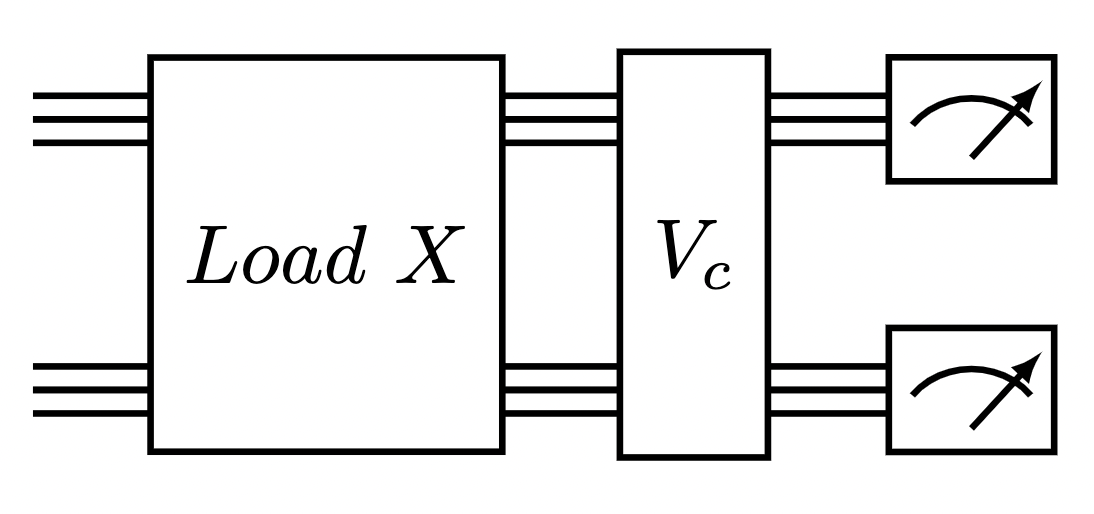}
    \caption{Quantum circuit to execute one attention layer of the Compound Transformer. We use a matrix data loader for $\mathbf{X}$ (equivalent to Fig.\ref{fig:dataloader_matrix}) and a quantum orthogonal layer for $\mathbf{V}_c$ applied on both registers.}
    \label{fig:compound_transformer}
\end{figure}

Until now, each step of the classical vision transformer has been reproduced closely by quantum linear algebraic procedures. The same quantum tools can also be used in a more natively quantum fashion, while retaining the spirit of the classical transformers, as shown in Fig.\ref{fig:compound_transformer}. 

At a high level, the compound transformer first loads all patches with the same weight applied each patch in superposition, and then apply an orthogonal layer that will at the same time extract the features from each patch and re-weight the patches so that in the end the output is computed as a weighted sum of the features extracted from all patches. This means that instead of calculating two separate weight matrices $\mathbf{V}$ and $\mathbf{W}$, one for feature extraction and one for weighting to generate  $\mathbf{y}_i =  \sum_j A'_{ij}\mathbf{V}\mathbf{x}_j$ individually, only one operation is used to generate all $\mathbf{y}_i$ directly from one circuit. Since a single quantum orthgonal layer is used to generate $\mathbf{Y}$, we switch to $\mathbf{V}_c$ to denote this orthogonal layer that applies the compound matrix as we explain below.

More precisely, the quantum circuit we use has two registers: the top one of size $n$ and the bottom one of size $d$. The full matrix $\mathbf{X} \in \R^{n \times d}$ is loaded into the circuit using the matrix data loader from Section \ref{sec:quantum_data_loaders} with $N=n+d$ qubits. This could correspond to the entire image, as every image can be split into $n$ patches of size $d$ each. Since the encoding basis over the two registers has more than one qubit in state 1, we are stepping out of the \emph{unary} basis framework. The correct basis to consider is of hamming weight $2$.
Note that, among the ${n+d}\choose{2}$ states with hamming weight 2, only $n \times d$ of them correspond to states with one 1 in the top qubits, and another 1 in the remaining bottom qubits.

Next, a quantum orthogonal layer $\mathbf{V}_c$ is applied on both registers at the same time. Note that this $\mathbf{V}_c$ is not the same as in the previous constructions, since now it is applied on a superposition of patches. As explained in Section \ref{sec:quantum_ortho_layers} and in \cite{subspace_states_qcware}, the resulting operation in this case is not a simple matrix-vector multiplication $\mathbf{VX}$. Instead of $\mathbf{V}$, the multiplication involves its $2^{nd}$-order compound matrix $\boldsymbol{\mathcal{V}}^{(2)}_c$ of dimension ${{n+d}\choose{2}} \times {{n+d}\choose{2}}$. Similarly, the vector multiplied is not simply $\mathbf{X}$ but a modified version of size ${n+d}\choose{2}$, obtained by padding the added dimensions with zeros.

The resulting state is $\ket{\mathbf{Y}} = \ket{\boldsymbol{\mathcal{V}}^{(2)}_cX}$, where $\boldsymbol{\mathcal{V}}^{(2)}_c$ is the $2^{nd}$-order compound matrix of the matrix $\mathbf{V}_c$, namely the matrix corresponding to the unitary of the quantum orthogonal layer in Fig.\ref{fig:compound_transformer} restricted to the unary basis. 
This state has dimension ${n+d}\choose{2}$, i.e. there 
are exactly two 1s in the $N=n+d$ qubits, but one can post-select for the part of the state where there is exactly one qubit in state 1 on the top register and the other 1 on the lower register. This way, $n\times d$ output states are generated. In other words, tomography is performed for a state of the form $\ket{\mathbf{Y}} = \frac{1}{\norm{\mathbf{Y}}}\sum_{i=1}^n\sum_{j=1}^d y_{ij}\ket{\mathbf{e}_j}\ket{\mathbf{e}_i}$ which is used to conclude that this quantum circuit produces transformed patches $(\mathbf{y}_1,\cdots, \mathbf{y}_n) \in \R^{n\times d}$. \revision{Note that in this context, the proposed tomography approach reconstructs vectors of a quadratic size, and not exponential, relative to the qubit count.} \revision{Furthermore, a significant fraction of the measurement shots might be discarded to narrow down to the desired $n \times d$ space as part of the the post-selection technique.}

To calculate the computational complexity of this circuit, we consider: the matrix data loader, detailed in Fig.\ref{fig:dataloader_matrix} which has depth of $\log n + 2n \log d$; the Quantum Orthogonal Layer applied on $n+d$ qubits, with a depth of $\log(n+d)$ and $(n+d)\log(n+d)$ trainable parameters if implemented using the butterfly circuit. Since this circuit uses exactly one matrix loader, the number of fixed parameters is $(n-1) + (2n-1)(d-1)$. 

In order to calculate the cost of performing the same operation on a classical computer, consider the equivalent operation of creating the compound matrix $\boldsymbol{\mathcal{V}}^{(2)}_c$ by first computing all determinants of the matrix and then performing a matrix-vector multiplication of dimension ${n+d}\choose{2}$, which takes $\mathcal{O}((n+d)^4)$ time. Performing this operation on a quantum computer can provide a polynomial speedup with respect to $n$. 
More generally, this compound matrix operation on an arbitrary input state of hamming weight $k$ is quite hard to perform classically, since all determinants must be computed, and a matrix-vector multiplication of size ${n+d}\choose{k}$ needs to be applied.

Overall, the compound transformer can replace both the Orthogonal Patch-wise Network (\ref{sec:ortho_patch_wise}) and the Quantum Transformer layer (\ref{sec:ortho_attention}) with one combined operation. The use of compound matrix multiplication makes this approach different from the classical transformers, while retaining some interesting properties with its classical counterpart: patches are weighted in their global context and gradients are shared through the determinants used to generate the compound matrix. 

The Compound Transformer operates in a similar spirit as the MLPMixer architecture presented in \cite{Tolstikhin2021MLPMixerAA}, which is a state-of-the-art architecture used for image classification tasks and  exchanges information between the different patches without using convolution or attention mechanisms. 

\section{Experiments}
\label{sec:experiments}

In order to benchmark the proposed methods, we applied them to a set of medical image classification tasks, using both simulations and quantum hardware experiments.  MedMNIST, a collection of 12 preprocessed, two-dimensional, open source medical image datasets from \cite{medmnist,medmnistv2},  annotated for classification tasks and benchmarking using a diverse set of classical techniques, is used to provide the complete training and validation data. 
\subsection{Simulation Setting}
\label{sec:simulation_setting}

Orthogonal Patch-wise Network from Section \ref{sec:ortho_patch_wise}, Orthogonal Transformer from Section \ref{sec:ortho_attention}, and Compound Transformer from Section \ref{sec:compound_transformers} were trained via simulation, along with two baseline methods. The first baseline is the Vision Transformer from \cite{dosovitskiy2020image},  which has been successfully applied to different image classification tasks and is described in detail in  \ref{sec:classical_transformer}. The second baseline is the Orthogonal Fully-Connected Neural Network (OrthoFNN), a quantum method without attention layer that has been previous trained on the RetinaMNIST dataset in \cite{Landman2022QuantumMF}. For each of the five architectures, one model was trained on each dataset of MedMNIST and validated using the same validation method as in \cite{medmnist,medmnistv2}. 

To ensure comparable evaluations between the five neural networks, similar architectures were implemented for all five.
The benchmark architectures all comprise of three parts: pre-processing, features extraction, and post-processing. The first part is classical and pre-processes the input image of size $28 \times 28$ by extracting $16$ patches ($n=16$) of size $7 \times 7$. We then map every patch to a $16$ dimensional feature space ($d = 16$) by using a fully connected neural network layer. 
This first feature extraction components is a single fully connected layer trained in conjunction to the rest of the architecture.
For the OrthoNN networks, used as our quantum baseline, one patch of size $16$ was extracted from the complete input image using a fully connected neural network layer of size $784 \times 16$. This fully connected layer is also trained in conjunction to the quantum circuits.
The second part of the common architecture transforms the extracted features by applying a sequence of $4$ attention layers on the extracted patches, which maintain the dimension of the layer. Moreover, the same gate layout, i.e. the butterfly circuit, is used for all circuits that compose the quantum layers. Finally, the last part of the neural network is classical, which linearly projects the extracted features and outputs the predicted label.

\subsection{Simulation Results}
\label{sec:simulation_results}

\begin{table*}[]
\centering
\resizebox{\linewidth}{!}{
\begin{tabular}{llcccccccccccc}
\hline
\multicolumn{2}{l}{\multirow{2}{*}{Network}} & \multicolumn{2}{c}{PathMNIST} & \multicolumn{2}{c}{ChestMNIST} & \multicolumn{2}{c}{DermaMNIST} & \multicolumn{2}{c}{OCTMNIST} & \multicolumn{2}{c}{PneumoniaMNIST} & \multicolumn{2}{c}{RetinaMNIST} \\
\multicolumn{2}{l}{} & AUC & ACC & AUC & ACC & AUC & ACC & AUC & ACC & AUC & ACC & AUC & ACC \\ \hline
\multicolumn{2}{l}{OrthoFNN (baseline)} & 0.939 & 0.643 & 0.701 & 0.947 & 0.883 & 0.719 & 0.819 & 0.516 & 0.950 & 0.864 & 0.731 & 0.548 \\
\multicolumn{2}{l}{OrthoPatchWise} & 0.953 & 0.713 & 0.692 & 0.947 & 0.898 & 0.730 & 0.861 & 0.554 & 0.945 & 0.867 & 0.739 & 0.560 \\ \hline
\multicolumn{2}{l}{VisionTransformer (baseline)} & 0.957 & 0.755 & \textbf{0.718} & \textbf{0.948} & 0.895 & 0.727 & \textbf{0.879} & \textbf{0.608} & \textbf{0.957} & \textbf{0.902} & 0.736 & 0.548 \\
\multicolumn{2}{l}{OrthoTransformer} & \textbf{0.964} & \textbf{0.774} & 0.703 & 0.947 & 0.891 & 0.719 & 0.875 & 0.606 & 0.947 & 0.885 & \textbf{0.745} & 0.542 \\
\multicolumn{2}{l}{CompoundTransformer} & 0.957 & 0.735 & 0.698 & 0.947 & \textbf{0.901} & \textbf{0.734} & 0.867 & 0.545 & 0.947 & 0.885 & 0.740 & \textbf{0.565} \\ \hline\hline
\multicolumn{2}{l}{\multirow{2}{*}{Network}} & \multicolumn{2}{c}{BreastMNIST} & \multicolumn{2}{c}{BloodMNIST} & \multicolumn{2}{c}{TissueMNIST} & \multicolumn{2}{c}{OrganAMNIST} & \multicolumn{2}{c}{OrganCMNIST} & \multicolumn{2}{c}{OrganSMNIST} \\
\multicolumn{2}{l}{} & AUC & ACC & AUC & ACC & AUC & ACC & AUC & ACC & AUC & ACC & AUC & ACC \\ \hline
\multicolumn{2}{l}{OrthoFNN (baseline)} & 0.815 & 0.821 & 0.972 & 0.820 & 0.819 & 0.513 & 0.916 & 0.636 & 0.923 & 0.672 & 0.875 & 0.481 \\ \multicolumn{2}{l}{OrthoPatchWise} & 0.830 & 0.827 & 0.984 & 0.866 & 0.845 & 0.549 & 0.973 & 0.786 & 0.976 & 0.805 & 0.941 & 0.640 \\
\hline
\multicolumn{2}{l}{VisionTransformer (baseline)} & 0.824 & 0.833 & \textbf{0.985} & \textbf{0.888} & \textbf{0.880} & \textbf{0.596} & 0.968 & 0.770 & 0.970 & 0.787 & 0.934 & 0.620 \\
\multicolumn{2}{l}{OrthoTransformer} & 0.770 & 0.744 & 0.982 & 0.860 & 0.856 & 0.557 & 0.968 & 0.763 & 0.973 & 0.785 & \textbf{0.946} & 0.635 \\
\multicolumn{2}{l}{CompoundTransformer} & \textbf{0.859} & \textbf{0.846} & \textbf{0.985} & 0.870 & 0.841 & 0.544 & \textbf{0.975} & \textbf{0.789} & \textbf{0.978} & \textbf{0.819} & 0.943 & \textbf{0.647} \\ \hline
\end{tabular}
}
\caption{Performance analysis using AUC and ACC on each test dataset of MedMNIST of our quantum architectures (Orthogonal PatchWise, Orthogonal Transformer and Compound Transformer) compared to the classical (Vision Transformer \cite{dosovitskiy2020image}) and quantum (Orthogonal FNN \cite{Landman2022QuantumMF}) baselines described in Section \ref{sec:experiments}.}
\label{table:simulation-results}
\end{table*}

A summary of the simulation results is shown in Table \ref{table:simulation-results} where the area under receiver operating characteristic (ROC) curve (AUC) and the accuracy (ACC) are reported as evaluation metrics. A full comparison with the classical benchmark provided by \cite{medmnist} is given in Appendix \ref{sec:extended-results}, Table \ref{table:simulation-results-all}.  

From Table \ref{table:simulation-results}, we observe that Vision Transformer, Orthogonal Transformer, and Compound Transformer architectures outperform the Orthogonal Fully-Connected and Orthogonal Patch-wise neural networks for all 12 tasks. This is likely due to the fact that the latter two architectures do not contain on any attention mechanism that exchange information across the patches, confirming the effectiveness of the attention mechanism to learn useful features from images. Second, Orthogonal Transformer and Compound Transformer, which implements non-trivial quantum attention mechanism, provide very competitive performances compared to the two benchmark methods and outperform the benchmark methods on 7 out of 12 MedMNIST datasets.

Moreover, comparisons can be made with regard to the number of trainable parameters used by each architecture.
Table \ref{table:resource_analysis} presents a resource analysis for the quantum circuits that were simulated per layer. E.g. the Compound Transformer requires 80 trainable parameters compared to the 512 ($2d^2$) required by the Classical Vision Transformer.
Note that this resource analysis focuses on the attention layer of each transformer network, and does not include parameters used for pre-processing, other parts found in the transformer layer, nor the single layer used in the final classification (Fig.\ref{fig:TransformerNetwork}), which are common to all simulated methods.

Overall, our quantum transformers have reached comparable levels of accuracy compared to the classical equivalent transformers, while using a smaller number of trainable parameters, providing confirmation of our theoretical predictions on a small scale. Circuit depth and number of distinct circuits used for each of the quantum transformers are also listed in Table \ref{table:resource_analysis} to match the theoretical resource analysis in Table \ref{table:quantum_transformer_comparison}. While the quantum transformers do have theoretical guarantee on the asymptotic run time for the attention mechanism compared to the classical transformer, this effect is hard to observe given the small data size. 
Summary of the hardware experiments listed in Table \ref{table:hardware results} shows very competitive levels of accuracy from the quantum transformers in comparison with the classical benchmarks. Details to be found in  \ref{sec:hardware_experiments}. 

\begin{table*}[]
\centering
\resizebox{\linewidth}{!}{
\begin{tabular}{|c|lc|cc|cc|}
\hline
\multirow{2}{*}{Model}                                            & \multicolumn{2}{l|}{Classical (JAX)} & \multicolumn{2}{c|}{IBM Simulator} & \multicolumn{2}{c|}{IBM Hardware} \\ \cline{2-7} 
                                                                  & AUC             & ACC                & AUC            & ACC               & AUC            & ACC              \\ \hline
Google AutoML (Best in \cite{medmnistv2})        & 0.750           & 53.10 \%           &   -             &   -                &     -           &    -              \\
VisionTransformer (classical benchmark)                       & 0.736           & 55.75 \%           &      -          &    -               &       -         &     -             \\ \hline
OrthoPatchWise (Pyramid Circuit)                                  & 0.738           & 56.50 \%           & 0.731          & 54.75 \%          & 0.727          & 51.75 \%         \\
Ortho Transformer (Pyramid Circuit)                               & 0.729           & 55.00 \%           & 0.715          & 55.00 \%          & 0.717          & 54.50 \%         \\
Ortho Transformer with Quantum Attention                                  & 0.749           & 56.50 \%           & 0.743          & 55.50 \%          & 0.746          & 55.00 \%         \\ \hline
CompoundTransformer (X Circuit)                                   & 0.729           & 56.50 \%           & 0.683          & 56.50 \%          & 0.666          & 45.75 \%         \\
CompoundTransformer ( \textbackslash~Circuit) & 0.716           & 55.75 \%           & 0.718          & 55.50 \%          & 0.704          & 49.00 \%         \\ \hline
\end{tabular}
}
\caption{Hardware Results for RetinaMNIST using various models. Classical (JAX): classical code run by JAX, equivalent to quantum operations. IBM Simulator: code compiled to run on actual IBM hardware and executed using their Aer Simulator. 
Note that ``\textbackslash~ Circuit" contains a single diagonal of trainable RBS gates. Details of the experiment are written in \ref{sec:hard_results_long}.} 
\label{table:hardware results}
\end{table*}

\begin{table*}[]
\centering
\resizebox{\linewidth}{!}{
\begin{tabular}{|c|c|c|c|c|c|}
\hline
\textbf{Model}      
& \textbf{Qubits} 
& \textbf{\begin{tabular}[c]{@{}c@{}}Number of Gates\\for Loaders\\ (Fixed Parameters)\end{tabular}} 
& \textbf{\begin{tabular}[c]{@{}c@{}}Number of Gates\\per Orthogonal Layer\\ (Trainable Parameters) \end{tabular}} 
& \textbf{Circuit Depth}
& \textbf{\begin{tabular}[c]{@{}c@{}}Number of \\Distinct Circuits\end{tabular}} \\ \hline
Orthogonal PatchWise   & 16   & 15   & 32   & 9  & 16  \\ \hline
Orthogonal Transformer & 16   & 45   & 64   & 9 \& 13  & 272     \\ \hline
Compound Transformer   & 32   & 480  & 80   & 150 & 1     \\ \hline
\end{tabular}
}
\caption{Resource analysis on a single attention layer used for the MedMNIST simulations (Section  \ref{sec:simulation_setting}).
\revision{From Table \ref{table:quantum_transformer_comparison}, it can be derived that the classical transformer requires 512 trainable parameters.}
Note that the \emph{Orthogonal Transformer} is using two different types of circuits per layer.}
\label{table:resource_analysis}
\end{table*}

\section{Conclusion}

In this work, three different quantum transformers are presented: Orthogonal Patchwise Transformer implements trivial attention mechanism; Orthogonal Transformer closely mimic the classical transformers; Compound Transformer steps away from the classical architecture with a quantum-native linear algebraic operation that cannot be efficiently done classically: multiplication of a vector with a higher-dimensional \emph{compound} matrix. Inside all these quantum transformers are the quantum orthogonal layers, which efficiently apply matrix multiplication on vectors encoded on specific quantum basis states. All circuits implementing orthogonal matrix multiplication can be trained using backpropagation detailed in \cite{Landman2022QuantumMF}.


\revisionnew{As shown in Table 2, the proposed quantum circuits offer a potential computational advantage in reducing the complexity of attention layers. This opens the possibility that quantum transformers may be able to match the performance of their classical counterparts, requiring fewer resources in terms of runtime and parameter count. On the other hand, while these initial results are promising, they are derived from a limited set of experiments and primarily offer a theoretical viewpoint. Practical realization of such advantages in quantum machine learning is heavily contingent upon future advancements in quantum hardware, for example in managing quantum noise, improving clock speed, and other critical factors. Therefore, these findings should be regarded as a promising yet preliminary step, necessitating further empirical validation using future quantum hardware.}

In addition to theoretical analysis, we performed extensive numerical simulations and quantum hardware experiments, which shows that our quantum circuits can classify the small MedMNIST images just as well as or at times better than the state-of-the-art classical methods (Table \ref{table:simulation-results}) while using fewer parameter, thereby showing potential of these quantum models to address over-fitting issues by using a smaller number of parameters. 

While the run time of the quantum fully connected layer and the quantum attention mechanism has been theoretically proven to be advantageous, this effect is hard to observe on the current quantum computers due to their limited size, high level of noise, and latency of cloud access.
From our hardware experiments, it can be observed that results from the current hardware become too noisy as soon as the number of qubits or the size of the quantum circuit increase. 

Overall, our results are encouraging and confirm the benefit of using trainable quantum circuits to perform efficient linear algebra operations. By carefully designing the quantum circuit to allow for much better control over the size of the Hilbert space that is explored by the model, we are able to provide models that are both expressive and trainable.

\newpage

\bibliographystyle{quantum}
\bibliography{references}

\clearpage

\appendix

\section{Vision Transformers}
\label{sec:classical_transformer}

Here, the details of a classical Vision Transformers introduced by \cite{dosovitskiy2020image} are outlined. Some slight changes in the architecture have been made to ease the correspondence with quantum circuits. We also introduce important notations that will be reused in the quantum methods.

The transformer network starts by decomposing an image into \emph{patches} and pre-processing the set of patches to map each one into a vector, as shown in Fig.\ref{fig:tranformerpatchdivision}.
The initial set of patches is enhanced with an extra vector of the same size as the patches, called \emph{class embedding}. This \emph{class embedding} vector is used at the end of the network, to feed into a fully connected layer that yields the output (see Fig.\ref{fig:TransformerNetwork}).
We also include one trainable vector called \emph{positional embedding}, which is added to each vector. At the end of this pre-processing step, we obtain the set of $n$ vectors of dimension $d$, denoted $\mathbf{x}_i$ to be used in the next steps. 

Next, feature extraction is performed using a transformer layer \cite{attentionisallyouneed, dosovitskiy2020image} which is repeated $L$ times, as shown in Fig.\ref{fig:tranformerlayer}. Within the transformer layer, we first apply layer normalisation over all patches $\mathbf{x}_i$, and then apply the attention mechanism detailed in Fig.\ref{fig:tranformerAttention}. After this part, we obtain a state to which we add the initial input vectors before normalisation in an operation called \emph{residual} layer, represented by the blue arrow in Fig.\ref{fig:tranformerlayer}, followed by another layer normalisation. After this, we apply a Multi Layer Perceptron (MLP), which consists of multiple fully connected linear layers for each vector that result in same-sized vectors. Again, we add the residual from just before the last layer normalisation, which is the output of one transformer layer.

After repeating the transformer layer $L$ times, we finally take the vector corresponding to the \emph{class embedding}, that is the vector corresponding to $\mathbf{x}_0$, in the final output and apply a fully connected layer of dimension ($d$ $\times$ number of classes) to provide the final classification result (see Fig.\ref{fig:TransformerNetwork}). It is important to observe here that we only use the first vector outcome in the final fully connected layer to do the classification (therefore the name \emph{class embedding}).

Looking inside the attention mechanism (see Fig.\ref{fig:tranformerAttention}), we start by using a fully connected linear layer with trainable weights $\mathbf{V}$ to calculate for each patch $\mathbf{x}_i$ the feature vector $\mathbf{V}\mathbf{x}_i$. Then to calculate the attention coefficients, we use another trainable weight matrix $\mathbf{W}$ and define the attention given by patch $\mathbf{x}_i$ to patch $\mathbf{x}_j$ as $\mathbf{x}_i^T\mathbf{W}\mathbf{x}_j$. Next, for each patch $\mathbf{x}_i$, we get the final extracted features as the weighted sum of all feature vectors $\mathbf{V}\mathbf{x}_j$ where the weights are the \emph{attention coefficients}. This is equivalent to performing a matrix multiplication with a matrix $\mathbf{A}$ defined by $A_{ij} = \mathbf{x}_i^T\mathbf{W}\mathbf{x}_j$. Note, in classical transformer architecture, a column-wise \emph{softmax} is applied to all $A_{ij}$ and attention coefficients $A'_{ij} = \text{softmax}_j(A_{ij})$ is used instead. Overall, the attention mechanism makes use of $2d^2$ trainable parameters, evenly divided between $\mathbf{V}$ and $\mathbf{W}$, each of size $d \times d$.

In fact, the above description is a slight variant from the original transformers proposed in \cite{attentionisallyouneed}, where the authors used two trainable matrices to obtain the attention coefficients instead of one ($\mathbf{W}$) in this work. This choice was made to simplify the quantum implementation but could be extended to the original proposal using the same quantum tools.

Computational complexity of classical attention mechanism depends mainly on the number of patches $n$ and their individual dimension $d$: the first patch-wise matrix multiplication with the matrix $\mathbf{V}\in\R^{d\times d}$ takes $\mathcal{O}(nd^2)$ steps, while the subsequent multiplication with the large matrix $\mathbf{A}'$ takes $\mathcal{O}(n^2d)$. Obtaining $\mathbf{A}'$ from $\mathbf{W}$ requires $\mathcal{O}(nd^2)$ steps as well. Overall, the complexity is $\mathcal{O}(nd^2 + n^2d)$. In classical deep learning literature, the emphasis is made on the second term, which is usually the most costly. Note that a recent proposal \cite{katharopoulos2020lineartransformers} proposes a different attention mechanism as a linear operation that only has a $\mathcal{O}(nd^2)$ computational complexity.

We compare the classical computational complexity with those of our quantum methods in Table \ref{table:quantum_transformer_comparison}. These running times have an real impact on both training and inference, as they measure how the time to perform each layer scales with the number and dimension of the patches.

\section{Quantum Tools (Extended)}
\label{sec:quantum_tools_extended}

\subsection{Quantum Data Loaders for Matrices}
\label{sec:quantum_data_loaders_long}

In order to perform a machine learning task with a quantum computer, classical data (a vector, a matrix) needs to be loaded into the quantum circuit. The technique we choose for this task is called \emph{amplitude encoding}, which uses the classical scalar component of the data as amplitudes of a quantum state made of $d$ qubits.  
In particular we build upon previous methods to define quantum data loaders for matrices, as shown in Fig.\ref{fig:dataloader_matrix}.

\begin{figure}[!h]
    \centering
    \includegraphics[width=0.45\textwidth]{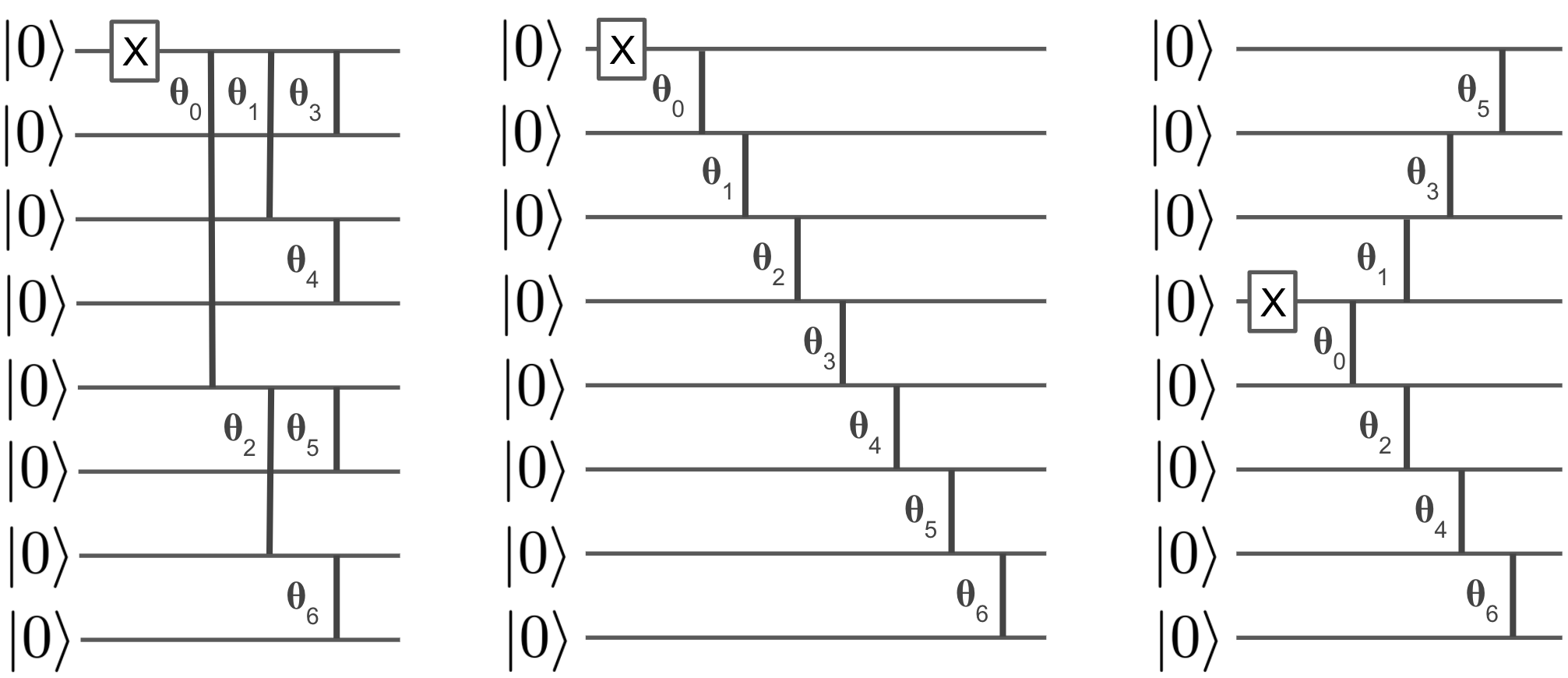}
    \caption{Three possible data loaders for $d$-dimensional vectors ($d=8$). From left to right: the parallel, diagonal, and semi-diagonal circuit have respectively a circuit depth of $\log(d)$, $d$, and $d/2$. The X gate represent the Pauli X gate, and the vertical lines represent RBS gates with tunable parameters.}
    \label{fig:dataloaders_vector}
\end{figure}

\cite{NearestCentroid2021} proposes three different circuits to load a vector $\mathbf{x}\in\R^d$ using $d-1$ gates for a circuit depth ranging from $\mathcal{O}(\log(d))$ to $\mathcal{O}(d)$ as desired (see Fig.\ref{fig:dataloaders_vector}). These data loaders use the \emph{unary} amplitude encoding, where a vector $\mathbf{x} = (x_1,\cdots,x_d)$ is loaded in the quantum state: $$\ket{\mathbf{x}} = \frac{1}{\norm{\mathbf{x}}}\sum_{i=1}^d x_i\ket{\mathbf{e}_i},$$ where $\ket{\mathbf{e}_i}$ is the quantum state with all qubits in $0$ except the $i^{th}$ one in state $1$ (e.g. $\ket{0\cdots010\cdots0}$). The circuit uses RBS gates: a parametrised two-qubit gate given by Eq.\ref{RBS}.

The $d-1$ parameters $\theta_i$ of the RBS gates are classically pre-computed to ensure that the output of the circuit is indeed $\ket{\mathbf{x}}$.

We require a data loader for matrices. Given a matrix $\mathbf{X} \in \R^{n\times d}$, instead of loading a flattened vector, rows $\mathbf{X}_i$ are loaded in superposition. As shown in Fig.\ref{fig:dataloader_matrix}, on the top qubit register, we first load the vector $(\norm{\mathbf{x} _1},\cdots,\norm{\mathbf{x}_n})$ made of the norms of each row, using a data loader for a vector and obtain a state $\frac{1}{\norm{\mathbf{X}}}\sum_{i=1}^n \norm{\mathbf{x} _i}\ket{\mathbf{e}_i}$. Then, on a lower register, we are sequentially loading each row $X_i \in \R^d$. To do so, we use vector data loaders and their adjoint, as well as CNOTs controlled on the $i^{th}$ qubit of the top register. The resulting state is a superposition of the form:

\begin{equation*}
    \ket{\mathbf{X}} = \frac{1}{\norm{\mathbf{X} }}\sum_{i=1}^n\sum_{j=1}^d X_{ij}\ket{\mathbf{e}_j}\ket{\mathbf{e}_i}
\end{equation*}

One immediate application of data loaders that construct amplitude encodings is the ability to perform fast inner product computation with quantum circuits. Applying the inverse data loader of $\mathbf{x}_i$ after the regular data loader of $\mathbf{x}_j$ effectively creates a state of the form $\braket{\mathbf{x}_i,\mathbf{x}_j}\ket{\mathbf{e}_1} + \ket{G}$ where $\ket{G}$ is a garbage state. The probability of measuring $\ket{\mathbf{e}_1}$, which is simply the probability of having a $1$ on the first qubit, is  $|\braket{\mathbf{x}_i,\mathbf{x}_j}|^2$. Techniques to retrieve the sign of the inner product have been developed in \cite{Landman2022QuantumMF}.

\subsection{Quantum Orthogonal Layers}
\label{sec:quantum_ortho_layers_long}

In this section, we outline the concept of quantum orthogonal layers used in neural networks, which generalises the work in \cite{Landman2022QuantumMF}. These layers correspond to parametrised circuits of $N$ qubits made of RBS gates. More generally, RBS gates preserve the number of ones and zeros in any basis state: if the input to a quantum orthogonal layer is a vector in unary amplitude encoding, the output will be another vector in unary amplitude encoding. Similarly, if the input quantum state is a superposition of only basis states of hamming weight $2$, so is the output quantum state. This output state is precisely the result of a matrix-vector product, where the matrix is the unitary matrix of the quantum orthogonal layer, restricted to the basis used. Therefore, for unary basis, we consider a $N \times N$ matrix $\mathbf{W}$ instead of the full $2^N \times 2^N$ unitary. Similarly for the basis of hamming weight two, we can restrict the unitary to a $N \choose 2$ $\times$ $N \choose 2$ matrix. Since the reduced matrix conserves its unitary property and has only real values, these are orthogonal matrices.
More generally, we can think of such hamming weight preserving circuits with $N$ qubits as block-diagonal unitaries that act separately on $N+1$ subspaces, where the $k$-th subspace is defined by all computational basis states with hamming weight equal to $k$. The dimension of these subspaces is equal to $N \choose k$.

There exist many possibilities for building a quantum orthogonal layer, each with different properties. The Pyramid circuit, proposed in \cite{Landman2022QuantumMF}, is composed of exactly $N(N-1)/2$ RBS gates. This circuit requires only adjacent qubit connectivity, which is the case for most superconducting qubit hardware. More precisely, the set of matrices that are equivalent to the quantum orthogonal layers with pyramidal layout is exactly the Special Orthogonal Group, made of orthogonal matrices with determinant equal to $+1$. We have showed that by adding a final Z gate on the last qubit would allow having orthogonal matrices with $-1$ determinant. The pyramid circuit is therefore very general and cover all the possible orthogonal matrices of size $N \times N$.

The two new types of quantum orthogonal layers we have introduced are the butterfly circuit (Fig.\ref{fig:butterfly}), and the $X$ circuit (Fig.\ref{fig:Xcircuit}) (Section \ref{sec:quantum_ortho_layers}).

There exists a method \cite{Landman2022QuantumMF} to compute the gradient of each parameter $\theta_i$ in order to update them. This backpropagation method for the pyramid circuit takes time $\mathcal{O}(N^2)$, corresponding to the number of gates, and provided a polynomial improvement in run time compared to the previously known orthogonal neural network training algorithms \cite{jia2019orthogonal}. The exact same method developed for the pyramid circuit can be used to perform quantum backpropagation on the new circuits introduced in this paper. The run time also corresponds to the number of gates, which is lower for the butterfly and $X$ circuits. See Table \ref{table:circuit_comparison} for full details on the comparison between the three types of circuits. \revision{In particular, when considering the butterfly layer, the complexity of the backpropagation method transitions from $\mathcal{O}(N^2)$ to $\mathcal{O}(N \log N)$.}

\section{Medical Image Classification via Quantum Transformers (Extended)}
\label{sec:experiments_extended}

\subsection{Datasets}

In order to benchmark our models, we used MedMNIST, a collection of 12 pre-processed, two-dimensional medical image open datasets \cite{medmnist,medmnistv2}. The collection has been standardised for classification tasks on 12 different imaging modalities, each with medical images of $28 \times 28$ pixels.  All three quantum transformers and two benchmark methods were trained and validated on all 12 MedMNIST datasets. For the hardware experiments, we focused on one dataset, RetinaMNIST. The MedMNIST dataset was chosen for our benchmarking efforts due to its accessible size for simulations of the quantum circuits and hardware experiments, while being representative of one important field of computer vision application: classification of medical images.

\subsection{Simulations}
\label{sec:simulations}

First, simulations of our models are performed on the 2D MedMNIST datasets and demonstrate that the proposed quantum attention architecture reaches accuracy comparable to and at times better than the various standard classical models. Next, the setting of our simulations are described and the results compared against those reported in the AutoML benchmark performed by the authors in \cite{medmnistv2}.  

\subsubsection{Simulation setting MedMNIST}
\label{simulationsMedmnistSettings}

The JAX package \cite{jax2018github} was used to  efficiently simulate the complete training procedure of the five benchmark architectures. The experimental hyperparameters used in \cite{medmnistv2} were replicated  for our benchmark: every model is trained using the cross-entropy loss with the Adam optimiser \cite{Kingma2015AdamAM} for $100$ epochs, with batch size of $32$ and a learning rate of $10^{-3}$ that is decayed by a factor of $0.1$ after $50$ and $75$ epochs.

The 5 different neural networks were trained over $3$ random seeds, and the best overall performance for each one of them was selected. The evaluation procedure is similar to the AutoML benchmark in \cite{medmnist,medmnistv2}, and the benchmark results are shown in Table \ref{table:simulation-results} where the area under receiver operating characteristic (ROC) curve (AUC) and the accuracy (ACC) are reported as evaluation metrics. A full comparison with the classical benchmark provided by \cite{medmnist} is given in (Appendix \ref{sec:extended-results}, Table \ref{table:simulation-results-all}).  

\subsubsection{Simulation results MedMNIST}

From Table \ref{table:simulation-results}, we observe that Quantum Orthogonal and Compound Transformer architectures outperform the Orthogonal Fully-Connected and Orthogonal Patch-wise neural networks most of the time. This may be due to the fact that the latter do not rely on any mechanism that exchange information across the patches. Second, all quantum neural networks provide very competitive performances compared to the AutoML benchmark and outperform their classical counterparts on 7 out of 12 MedMNIST datasets.

Moreover, comparisons can be made with regard to the number of parameters used by each architecture, in particular for feature extraction.
Table \ref{table:resource_analysis} presents a resource analysis for the quantum circuits that were simulated, per layer. It includes the number of qubits, the number of gates with trainable parameters, and the number of gates with fixed parameters used for loading the data. The table shows that our quantum architectures have a small number of trainable parameters per layer. The global count for each quantum method is as follows.
\begin{itemize}
    \item Orthogonal Patch-wise Neural Network: $32$ parameters per circuit, 16 circuits per layer which use the same 16 parameters, and 4 layers, for a total of $128$ trainable parameters.
    \item Quantum Orthogonal Transformer: $32$ parameters per circuit, $17$ circuits which use the same 16 parameters and another $289$ circuits which use another set of 16 parameters per layer, and 4 layers, for a total of $256$ trainable parameters.
    \item Compound Transformer: $80$ parameters per circuit, 1 circuit per layer, and 4 layers, for a total of $320$ trainable parameters.
\end{itemize}  
These numbers are to be compared with the number of trainable parameters in the classical Vision Transformer that is used as a baseline. As stated in Section \ref{sec:classical_transformer}, each classical attention layer requires $2d^2$ trainable parameters, which in the simulations performed here corresponds to 512. Note again this resource analysis focuses on the attention layer of the each transformer network, and does not include parameters used for the preprocessing of the images (see Section \ref{simulationsMedmnistSettings}), as part of other transformer layers (Fig.\ref{fig:tranformerlayer}), and for the single layer used in the final classification (Fig.\ref{fig:TransformerNetwork}), which are common in all cases. 

More generally, performance of other classical neural network models provided by the authors of MedMNIST is compared to our approaches in Table \ref{table:simulation-results-all} found in the Appendix. Some of these classical neural networks reach somewhat better levels of accuracy, but are known to use an extremely large number of parameters. For instance, the smallest reported residual network has approximately a total number of $10^7$ parameters, and the automated machine learning algorithms train numerous different architectures in order to reach that performance.

Based on the results of the simulations in this section, quantum transformers are able to train across a number different of classification tasks, deliver performances that are highly competitive and sometimes better than the equivalent classical methods.

\subsection{Quantum Hardware Experiments}
\label{sec:hardware_experiments}

Quantum hardware experiments were performed on one specific dataset: RetinaMNIST. It has $1080$ images for training, $120$ images for validation, and $400$ images for testing. Each image contains $28\times28$ RGB pixels. Each image is classified into 1 of 5 classes (ordinal regression).

\subsubsection{Hardware Description}
\begin{figure}[!h]
    \centering
    \includegraphics[width=0.3\textwidth]{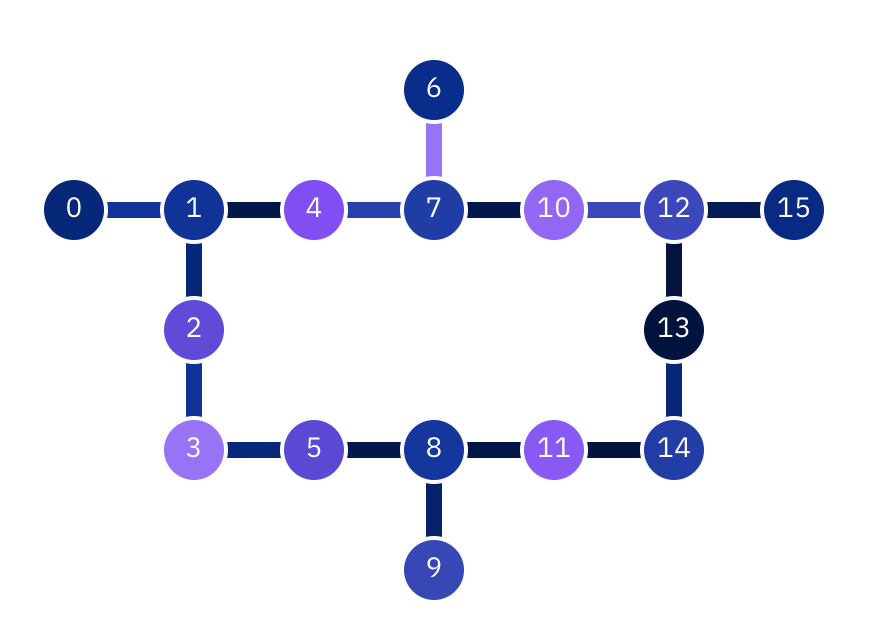}
    \caption{Connectivity of the 16-qubit ibmq\_guadalupe quantum computer.}
    \label{guadalupe}
\end{figure}

The hardware demonstration was performed on two different superconducting quantum computers provided by IBM, with the smaller experiments performed on the 16-qubit {\em ibmq\_guadalupe} machine (see Fig.\ref{guadalupe}) and the larger ones on the 27-qubit {\em ibm\_hanoi} machine.
Results are reported here from experiments with four, five and six qubits; experiments with higher numbers of qubits, which entails higher numbers of gates and depth, did not produce meaningful results.

\begin{figure}[!h]
    \centering
    \includegraphics[width=0.45\textwidth]{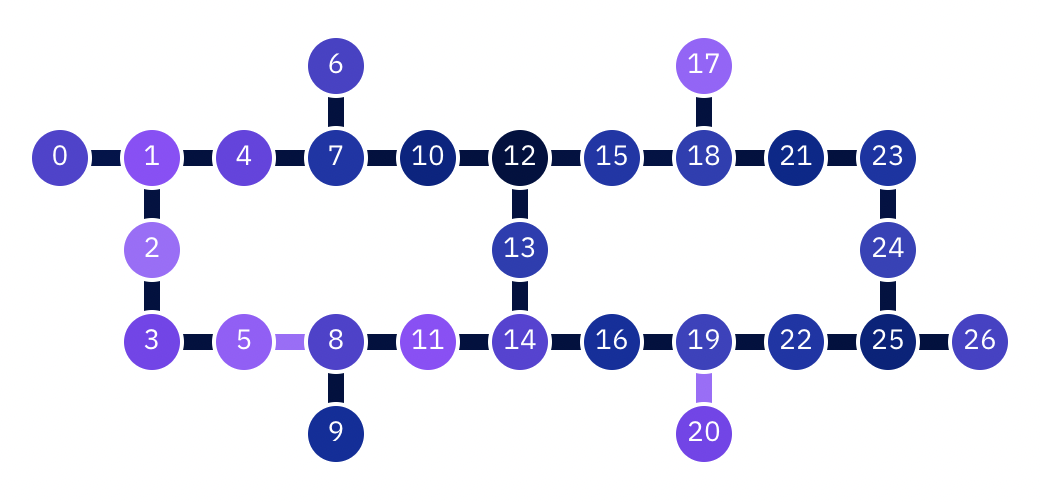}
    \caption{Connectivity of the 27-qubit ibm\_hanoi quantum computer.}
    \label{hanoi}
\end{figure}

Note that the main sources of noise are the device noise and the finite sampling noise. In general, noise is undesirable during computations. In the case of a neural network, however, noise may not be as troublesome: noise can help escape local minima \cite{noiseNN2017}, or act as data augmentation to avoid over-fitting. In classical deep learning, noise is sometimes artificially added for these purposes \cite{ying2019overview}. Despite this, when the noise is too large, we also see a drop in the accuracy.

\subsubsection{Hardware Results}
\label{sec:hard_results_long}

Hardware experiments were performed with four, five and six qubits to push the limits of the current hardware, in terms of both the number of qubits and circuit depth. Three quantum proposals were run: the Orthogonal Patch-wise network (from Section \ref{sec:ortho_patch_wise}), the Quantum Orthogonal transformers (from Sections \ref{sec:quantum_transformers} and \ref{sec:quantum_direct_attention}) and finally the Quantum Compound Transformer (from Section \ref{sec:compound_transformers}). 

Each quantum model was trained using a JAX-based simulator, and inference was performed on the entire test dataset of $400$ images of the RetinaMNIST on the IBM quantum computers. \revision{Regarding the experimental setting on real hardware, the number of shots for the compound setup using $6$ qubits was maximized to $32.000$. For other configurations using 4 qubits, $10.000$ shots were used.} 

The first model, the Orthogonal Patch-wise neural network, was trained using 16 patches per image, 4 features per patch, and one $4\times4$ orthogonal layer, using a 4-qubit pyramid as the orthogonal layer. The experiment used 16 different quantum circuits of 9 RBS gates per circuit per image. The result was compared with an equivalent classical (non-orthogonal) patch-wise neural network, and a small advantage in accuracy for the quantum native method could be reported.

The second model, the Quantum Orthogonal Transformer, used $4$ patches per image, 4 features per patch, and an attention mechanism with one $4\times 4$ orthogonal layer and trainable attention coefficients. 4-qubit pyramids were used as orthogonal layers. The experiment used 25 different quantum circuits of 12 RBS gates per circuit per image and 15 different quantum circuits of 9 RBS gates per circuit per image.

The third set of experiments ran the Orthogonal Transformer with the quantum attention mechanism. 
We used $4$ patches per image, 4 features per patch, and a quantum attention mechanism that paid attention to only the neighbouring patch, thereby using a $5$-qubit quantum circuit with the $X$ as the orthogonal layer. The experiment used 12 different quantum circuits of 14 RBS gates and 2 $CNOT$s per circuit per image.

The last two quantum proposals were compared with a classical transformer network with a similar architecture and demonstrated similar level of accuracy. 

Finally, the fourth experiment was performed on the \textit{ibmq\_hanoi} machine with 6 qubits, with the Compound Transformer, using 4 patches per image, 4 features per patch, and one orthogonal layer using the $X$ layout. The hardware results were quite noisy with the X layer, therefore the same experiments were performed with a further-reduced orthogonal layer named the ``\textbackslash Circuit": half of a X Circuit (Fig.\ref{fig:Xcircuit}) where only one diagonal of RBS gates is kept, and which reduced the noise in the outcomes. The experiment used 2 different quantum circuits of 18 RBS gates and 3 $CNOT$s per circuit per image.

Note that with the restriction to states with a fixed hamming weight, strong error mitigation techniques become available. Indeed, as we expect to obtain only quantum superpositions of unary states or states with hamming weight $2$ in the case of Compound Transformers, at every layer, every measurement can be processed to discard the ones that have a different hamming weight \emph{i.e.} states with more than one (or two) qubit in state $\ket{1}$. This error mitigation procedure can be applied efficiently to the results of a hardware demonstration, and has been used in the results presented in this paper. 

The conclusion from the hardware experiments is that all quantum proposals achieve state-of-the-art test accuracy, comparable to classical networks.
\revision{Looking at the simulation experiments (details found in Table \ref{table:simulation-results}), the compound transformer occasionally achieves superior performance compared to classical transformer. Note that achieving such a compound implementation in a classical setting incurs a polynomial overhead.} 

\section{Extended Performance Analysis}\label{sec:extended-results}

\begin{table*}[h!]
  \centering
  \resizebox{\linewidth}{!}{
    \begin{tabular}{llcccccccccccc}
      \hline
      \multicolumn{2}{l}{\multirow{2}{*}{Network}} & \multicolumn{2}{c}{PathMNIST} & \multicolumn{2}{c}{ChestMNIST} & \multicolumn{2}{c}{DermaMNIST} & \multicolumn{2}{c}{OCTMNIST} & \multicolumn{2}{c}{PneumoniaMNIST} & \multicolumn{2}{c}{RetinaMNIST} \\
      \multicolumn{2}{l}{}                         & AUC   & ACC   & AUC   & ACC   & AUC   & ACC   & AUC   & ACC   & AUC   & ACC   & AUC   & ACC   \\ \hline
      \multicolumn{2}{l}{\textit{ResNet-18 (28)}}  & 0.983 & 0.907 & 0.768 & 0.947 & 0.917 & 0.735 & 0.943 & 0.743 & 0.944 & 0.854 & 0.717 & 0.524 \\
      \multicolumn{2}{l}{\textit{ResNet-18 (224)}} & 0.989 & 0.909 & 0.773 & 0.947 & 0.920 & 0.754 & 0.958 & 0.763 & 0.956 & 0.864 & 0.710 & 0.493 \\
      \multicolumn{2}{l}{\textit{ResNet-50 (28)}}  & 0.990 & 0.911 & 0.769 & 0.947 & 0.913 & 0.735 & 0.952 & 0.762 & 0.948 & 0.854 & 0.726 & 0.528 \\
      \multicolumn{2}{l}{\textit{ResNet-50 (224)}} & 0.989 & 0.892 & 0.773 & 0.948 & 0.912 & 0.731 & 0.958 & 0.776 & 0.962 & 0.884 & 0.716 & 0.511 \\
      \multicolumn{2}{l}{\textit{auto-sklearn}}    & 0.934 & 0.716 & 0.649 & 0.779 & 0.902 & 0.719 & 0.887 & 0.601 & 0.942 & 0.855 & 0.690 & 0.515 \\
      \multicolumn{2}{l}{\textit{auto-keras}}      & 0.959 & 0.834 & 0.742 & 0.937 & 0.915 & 0.749 & 0.955 & 0.763 & 0.947 & 0.878 & 0.719 & 0.503 \\
      \multicolumn{2}{l}{\textit{auto-ml}}         & 0.944 & 0.728 & 0.914 & 0.948 & 0.914 & 0.768 & 0.963 & 0.771 & 0.991 & 0.946 & 0.750 & 0.531 \\\hline
      \multicolumn{2}{l}{VisionTransformer}        & 0.957 & 0.755 & 0.718 & 0.947 & 0.895 & 0.727 & 0.923 & 0.830 & 0.957 & 0.902 & 0.749 & 0.562 \\
      \multicolumn{2}{l}{OrthoFNN}                 & 0.939 & 0.643 & 0.701 & 0.947 & 0.883 & 0.719 & 0.819 & 0.516 & 0.950 & 0.864 & 0.731 & 0.548 \\ \hline
      \multicolumn{2}{l}{OrthoPatchWise}           & 0.953 & 0.713 & 0.692 & 0.947 & 0.898 & 0.730 & 0.861 & 0.554 & 0.945 & 0.867 & 0.739 & 0.560 \\
      \multicolumn{2}{l}{OrthoTransformer}         & 0.964 & 0.774 & 0.703 & 0.947 & 0.891 & 0.719 & 0.875 & 0.606 & 0.947 & 0.885 & 0.745 & 0.542 \\
      \multicolumn{2}{l}{CompoundTransformer}      & 0.957 & 0.735 & 0.698 & 0.947 & 0.901 & 0.734 & 0.867 & 0.545 & 0.947 & 0.885 & 0.740 & 0.565 \\ \hline\hline
      \multicolumn{2}{l}{\multirow{2}{*}{Network}} & \multicolumn{2}{c}{BreastMNIST} & \multicolumn{2}{c}{BloodMNIST} & \multicolumn{2}{c}{TissueMNIST} & \multicolumn{2}{c}{OrganAMNIST} & \multicolumn{2}{c}{OrganCMNIST} & \multicolumn{2}{c}{OrganSMNIST} \\
      \multicolumn{2}{l}{}                         & AUC   & ACC   & AUC   & ACC   & AUC   & ACC   & AUC   & ACC   & AUC   & ACC   & AUC   & ACC   \\ \hline
      \multicolumn{2}{l}{\textit{ResNet-18 (28)}}  & 0.901 & 0.863 & 0.998 & 0.958 & 0.930 & 0.676 & 0.997 & 0.935 & 0.992 & 0.900 & 0.972 & 0.782 \\
      \multicolumn{2}{l}{\textit{ResNet-18 (224)}} & 0.891 & 0.833 & 0.998 & 0.963 & 0.933 & 0.681 & 0.998 & 0.951 & 0.994 & 0.920 & 0.974 & 0.778 \\
      \multicolumn{2}{l}{\textit{ResNet-50 (28)}}  & 0.857 & 0.812 & 0.997 & 0.956 & 0.931 & 0.680 & 0.997 & 0.935 & 0.992 & 0.905 & 0.972 & 0.770 \\
      \multicolumn{2}{l}{\textit{ResNet-50 (224)}} & 0.866 & 0.842 & 0.997 & 0.950 & 0.932 & 0.680 & 0.998 & 0.947 & 0.993 & 0.911 & 0.975 & 0.785 \\
      \multicolumn{2}{l}{\textit{auto-sklearn}}    & 0.836 & 0.803 & 0.984 & 0.878 & 0.828 & 0.532 & 0.963 & 0.762 & 0.976 & 0.829 & 0.945 & 0.672 \\
      \multicolumn{2}{l}{\textit{auto-keras}}      & 0.871 & 0.831 & 0.998 & 0.961 & 0.941 & 0.703 & 0.994 & 0.905 & 0.990 & 0.879 & 0.974 & 0.813 \\
      \multicolumn{2}{l}{\textit{auto-ml}}         & 0.919 & 0.861 & 0.998 & 0.966 & 0.924 & 0.673 & 0.990 & 0.886 & 0.988 & 0.877 & 0.964 & 0.749 \\\hline
      \multicolumn{2}{l}{VisionTransformer}        & 0.824 & 0.833 & 0.985 & 0.888 & 0.880 & 0.596 & 0.968 & 0.770 & 0.970 & 0.787 & 0.934 & 0.620 \\
      \multicolumn{2}{l}{OrthoFNN}                 & 0.815 & 0.821 & 0.972 & 0.820 & 0.819 & 0.513 & 0.916 & 0.636 & 0.923 & 0.672 & 0.875 & 0.481 \\ \hline
      \multicolumn{2}{l}{OrthoPatchWise}           & 0.830 & 0.827 & 0.984 & 0.866 & 0.845 & 0.549 & 0.973 & 0.786 & 0.976 & 0.805 & 0.941 & 0.640 \\
      \multicolumn{2}{l}{OrthoTransformer}         & 0.770 & 0.744 & 0.982 & 0.860 & 0.856 & 0.557 & 0.968 & 0.763 & 0.973 & 0.785 & 0.946 & 0.635 \\
      \multicolumn{2}{l}{CompoundTransformer}      & 0.859 & 0.846 & 0.985 & 0.870 & 0.841 & 0.544 & 0.975 & 0.789 & 0.978 & 0.819 & 0.943 & 0.647 \\ \hline
    \end{tabular}
  }
  \caption{Extended Performance Analysis in terms of AUC and ACC on each test dataset of MedMNIST. \revision{The numbers (28) and (224) next to ResNet denote the input image resolutions of 28x28 and 224x224, respectively, with the latter being a larger version of ResNet with more parameters. These benchmarks are based on the results reported in \cite{medmnistv2}.}}
  \label{table:simulation-results-all}
\end{table*}

We add our results to the already existing results on the MedMNIST \cite{medmnistv2} datasets in the Table \ref{table:simulation-results-all} below.

\end{document}